\documentclass{aastex62}

\usepackage{graphicx}	% Including figure files
\usepackage{amsmath}	% Advanced maths commands
\usepackage{amssymb}	% Extra maths symbols

\received{January 1, 2018}
\revised{January 7, 2018}
\accepted{\today}
\submitjournal{ApJS}

\newcommand{\teff}{${T}_{\rm eff}$}
\newcommand{\logg}{$\log{g}$}
\newcommand{\feh}{{\rm [Fe/H]}}

\newcommand{\mh}{[M/H]}

\newcommand{\am}{[$\alpha$/M]}
\newcommand{\cm}{[C/M]}
\newcommand{\nm}{[N/M]}

\newcommand{\cannon}{\emph{The Cannon}}

\newcommand{\slam}{{SLAM}}
\newcommand{\lamost}{LAMOST}

\newcommand{\apogee}{APOGEE}
\newcommand{\aspcap}{ASPCAP}

\newcommand{\bz}[1]{\emph{\textcolor[rgb]{1.00,0.00,0.00}{authors: [#1]}}}
\renewcommand{\bz}[1]{#1}

\shorttitle{Stellar LAbel Machine (\slam)}
\shortauthors{Zhang et al.}
\begin{document}

\title{Deriving the stellar labels of LAMOST spectra with Stellar LAbel Machine (\slam)}

\correspondingauthor{Bo Zhang}
\email{bozhang@nao.cas.cn}

\author[0000-0002-6434-7201]{Bo Zhang}
\affil{Key Laboratory of Optical Astronomy, National Astronomical Observatories, Chinese Academy of Sciences, Beijing 100101,  People's Republic of China}
\affil{University of Chinese Academy of Sciences, Beijing 100049, People's Republic of China}

\author[0000-0002-1802-6917]{Chao Liu}
\affil{Key Laboratory of Optical Astronomy, National Astronomical Observatories, Chinese Academy of Sciences, Beijing 100101,  People's Republic of China}

\author[0000-0001-9073-9914]{Li-Cai Deng}
\affil{Key Laboratory of Optical Astronomy, National Astronomical Observatories, Chinese Academy of Sciences, Beijing 100101,  People's Republic of China}



\begin{abstract}
The LAMOST survey has provided 9 million spectra in its Data Release 5 (DR5) at $R\sim1800$.
Extracting precise stellar labels is crucial for such a large sample.
In this paper, we report the implementation of the Stellar LAbel Machine (SLAM), which is a data-driven method based on Support Vector Regression (SVR), a robust non-linear regression technique.
Thanks to the capability to model highly non-linear problems with SVR, SLAM generally can derive stellar labels over a wide range of spectral types. This gives it a unique capability compared to other popular data-driven methods. 
To illustrate this capability, we test the performance of SLAM on stars ranging from $T_{\rm eff}\sim4000$ to $\sim8000$\,K trained on LAMOST spectra and stellar labels.
At $g$-band signal-to-noise ratio (SNR$_g$) higher than 100, the random uncertainties of $T_{\rm eff}$, $\log g$ and [Fe/H] are 50\,K, 0.09\,dex, and 0.07\,dex, respectively.
We then set up another SLAM model trained by APOGEE and LAMOST common stars to demonstrate its capability of dealing with high dimensional problems. The spectra are from LAMOST DR5 and the stellar labels of the training set are from  APOGEE DR15, including $T_{\rm eff}$, $\log g$, [M/H], [$\alpha$/M], [C/M], and [N/M]. The cross-validated scatters at SNR$_g\sim100$ are 49\,K, 0.10 \,dex, 0.037\,dex, 0.026\,dex, 0.058\,dex, and 0.106\,dex for these parameters, respectively. This performance is at the same level as other up-to-date data-driven models.
%We, therefore, conclude that SLAM is ready for deriving multi-dimensional stellar labels in high precision over a larger range of spectral types.
As a byproduct, we also provide the latest catalog of $\sim1$ million LAMOST DR5 K giant stars with \slam-predicted stellar labels in this work.

\end{abstract}

\keywords{methods: data analysis --- methods: statistical --- stars: abundances --- stars: fundamental parameters --- catalogs --- surveys}


\section{Introduction}
As large spectroscopic surveys, e.g., SDSS/SEGUE \citep{Beers2006}, RAVE \citep{Steinmetz2006}, SDSS/APOGEE \citep{Majewski2012},  LAMOST \citep{2012RAA....12..735D}, Gaia-ESO \citep{Gilmore2012}, and GALAH \citep{Freeman2012} proceed, deriving the stellar labels (or stellar parameters) is of extreme importance. In particular, such large surveys often observe stars covering a large range of spectral types.
LAMOST, for instance, has observed stars from O type to M type \citep{Liu2019, Zhong2019}. This requires that the stellar label estimator must be able to deal with stellar samples over a large range of spectral types.

Stellar labels are usually determined by comparing an observed spectrum to a stellar spectral library (either a pre-computed synthetic or empirical stellar spectral library).
\bz{
The idea of data-driven methods \citep[\cannon, ][]{Ness2015} is proposed for its capability to set up the mappings from stellar labels to spectra with a training set and use them to predict stellar labels for the observed spectra.}
It is not only proved competitive to \aspcap\ in APOGEE case \citep{Ness2015} but also demonstrated the capability in predicting stellar labels from the low-resolution spectra of LAMOST K giant stars \citep{Ho2017a, Ho2017b}.
\bz{
\cite{Casey2017} and \cite{2018MNRAS.478.4513B} extended the application of \cannon\ to main sequence stars in the analysis of the RAVE and GALAH data, respectively.
Based on quadratic models, improvements such as regularization have been made to make \cannon\ capable to predict stellar labels  more precisely \citep{Casey2016, Casey2017}.}

In the training stage, with a training set, a \cannon-like method uses regression methods to build a generative model of the spectral flux at a given wavelength as a function of stellar labels, i.e.,
\begin{equation}
F(\lambda)=f_\lambda({T}_{\rm eff}, \log{g},  {\rm [X/H]}, ...),
\end{equation}
where $F(\lambda)$ is the normalized spectral flux at wavelength $\lambda$, $f_\lambda$ is the assumed form of spectral flux at $\lambda$ and \teff, \logg\ and ${\rm [X/H]}$ are the stellar effective temperature, surface gravity and elemental abundances, respectively.
\bz{In \cite{Ness2015}, $f_\lambda$ is adopted as a quadratic function whose coefficients are optimized in the training process in order to well fit the training set.}
A more general case is discussed by \cite{2016ApJ...826L..25R}.
In the \bz{test step}, the stellar labels are determined by operating the generative model to search for a model spectrum that best fits the observed one.

\begin{figure*}
\plotone{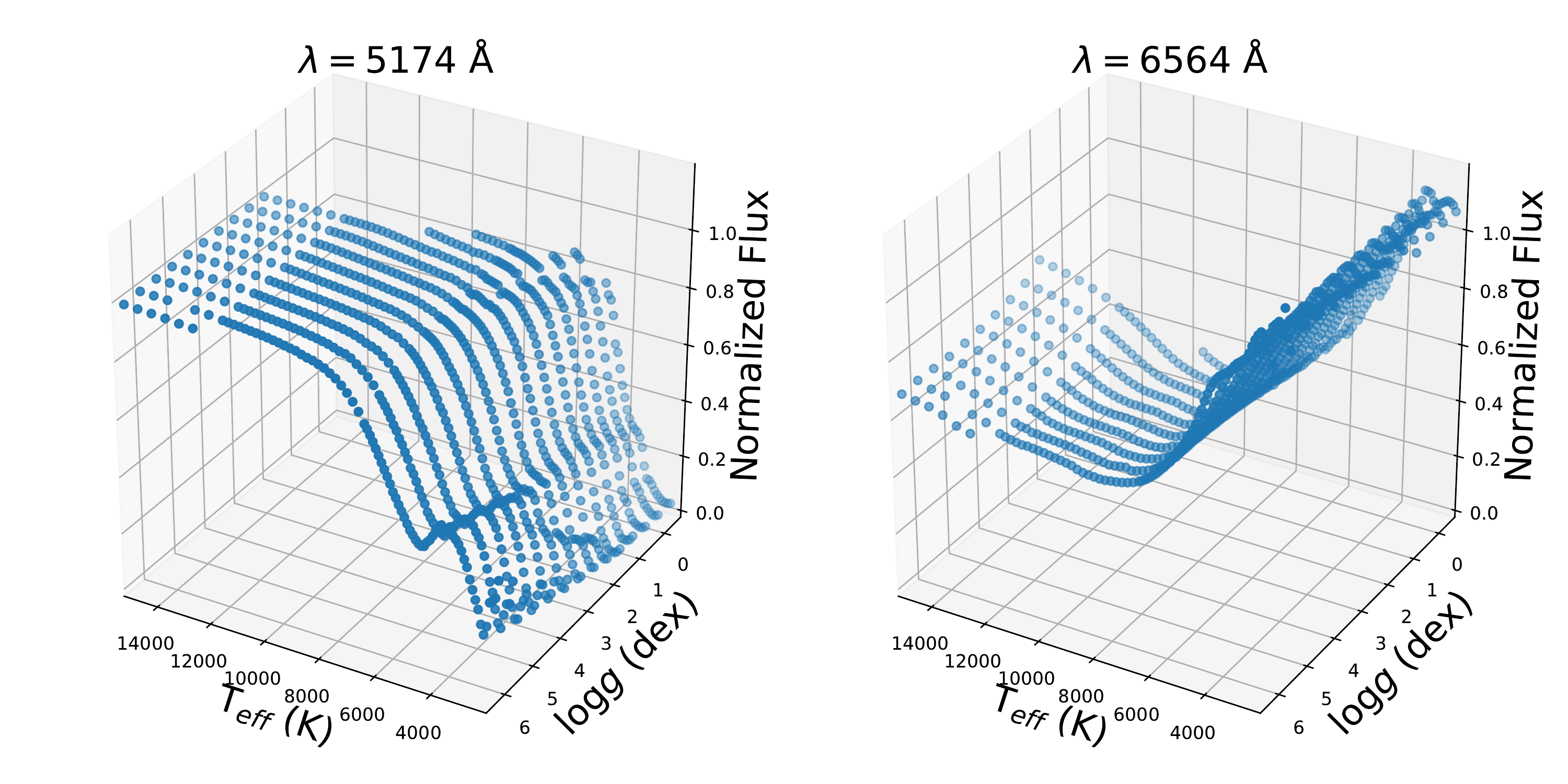}
\caption{Examples of how spectral flux changes with two primary stellar labels, i.e., \teff\ and \logg, at two fixed wavelengths. 
Blue dots in left and right panels are the flux values of normalized PHOENIX spectra with \feh$=0.0$ at $\lambda=5174$ \AA (Mg $b$) and 6564 \AA  ($\rm{H}\alpha$), respectively.
%Red dots show the flux predicted using SVR at the corresponding grid.
Since these two pixels are around the spectral lines which are extremely important in deriving stellar labels, a model's fitting performance for such data is crucial for stellar label prediction.
\label{fig_phoenix_pixel}}
\end{figure*}

\bz{The idea of data-driven methods is important. However, a better form of $f_\lambda$ is needed when modeling spectra that cover a wide range of spectral types.} For instance, at around some strong atomic lines, fluxes can dramatically change in highly non-linear ways with \teff\ and \logg.
In the left/right panel of Figure \ref{fig_phoenix_pixel}, we show the trends of normalized synthetic fluxes from PHOENIX library \citep{Husser2013} at around ${\rm Mg}\ b$ / ${\rm H}\alpha$. 
It is clearly seen that a quadratic function is no longer sufficient to associate the stellar labels with spectral fluxes when \teff\ changes from 3000 to 15000\,K.
This is also shown in \cite{2019ApJ...879...69T}.
%
%% ======= 2019-08-04 ========
%

One possible solution, the Payne \citep{Ting2017, 2019ApJ...879...69T}, is based on Neural-Networks (NN).
It is fascinating because in the training stage the cost function of the Payne is regularized by a synthetic gradient.
However, an NN-based method may suffer from the 'when-to-stop' problem because the learning curve would not tell one when the NN is optimized (neither \textit{over-fitting} nor \textit{under-fitting}). As a consequence, the optimization of these kinds of methods depend on expertise and experience of the users.

%Compare with:
%\begin{enumerate}
%\item Generative NN (wang rui)
%\item KPCA
%\end{enumerate}
%If we insist on using parametric models, a possible way is to extend the polynomial model to a higher order, which brings in more degrees of freedom in order to fit the complicated training data and therefore predict more precisely.
%However, such a parametric model may still lack of variance and introduce in some unreasonable systematics in predictions. For example, if we select a 10th order polynomial model, questions of why we should use 10th order polynomial occur immediately. Why don't we use a 11th order polynomial? What if we use a even higher order model? Would that be more precise or it's just over-fitting data?

%Once we decide to use non-parametric models, there are a bunch of tools we can turn to, such as various kinds of splines, lots of tree-based methods and neural-networks.
\bz{
The Support Vector Regression \citep[SVR, ][]{ss2004, Chang2011}, which is well-known for its robustness in modeling noisy data, is not a newcomer in the field of spectral data analysis.
It is often used to build the mapping from stellar spectra to stellar labels in previous works \cite{2014RAA....14..423L, 2014ApJ...790..105L, 2015MNRAS.447..256B, 2015MNRAS.452.1394L}.
In this paper, following the idea of data-driven approaches, we present an alternative implementation, the Stellar LAbel Machine (SLAM), by using SVR to build a generative model of spectra, which automatically adjusts the \textit{model complexity} according to data and robustly extract as much information as possible from stellar spectra.
}
Section \ref{sec:slam} gives a brief description of \slam\, and Section \ref{sec:test} assesses the performance of \slam\ using the LAMOST DR5.
%In section 4 we use \slam\ to replicate the stellar labels from APOGEE DR13 spectra.
In Section \ref{sec:apogee}, we predict stellar labels for \lamost\ DR5 K giant stars using \slam\ with \apogee\ DR15 stellar labels as the training data. Then we present the resulting catalog of more than a million red giant stars with precise stellar labels.
\bz{
We discuss the advantages and disadvantages of \slam\ in Section \ref{sec:discussion} and present the coefficients of dependence (CODs) in Section \ref{sec:cods}.
At last, we draw the conclusions of this paper in Section \ref{sec:conclusion}.
}
%It is able to tackle arbitrary dimensional parameter space in principle although computationally expensive compared to simple polynomial model.
%Here we report our implementation of a new method named Stellar LAbel Machine (\slam), which is based on SVR therefore named after Support Vector Machine (SVM).
%Since SVR is a non-parametric regression method, \slam\ is able to overcome the weakness of \cannon\ described above.
%The obvious major advantages of \slam\ consist of
%\begin{enumerate}
%\item the essence of a non-parametric way of modeling pixels based on Support Vector Regression (SVR),
%\item the ability to tackle various kinds of complicated training data,
%\item and integration of MCMC which enables users to derive the PDFs of stellar labels therefore get more precise error estimations.
%\end{enumerate}

\section{The construction of \slam} \label{sec:slam}

In principle, \slam\ consists of 3 steps.
\begin{enumerate}
\item The first step is data pre-processing. This includes spectra normalization and training data standardization, e.g., re-scale both stellar labels and spectral fluxes so that their mean is 0 and variance is 1.
\item The second step is to train SVR model at each wavelength pixel using the training set.
\item And the last is to predict stellar labels for observed spectra.
\end{enumerate}
The details are described in the below.

\subsection{Pre-processing}
This step is to map all the normalized spectral fluxes and the stellar labels of the training set in standardized space (with zero mean and unity variance). It is necessary for most machine learning methods, including SVR, to avoid issues due to the different scales in different dimensions of the input data.

\bz{After correcting its radial velocity (RV), each stellar spectrum in the training set is normalized by dividing its pseudo-continuum.
In SLAM, we firstly use a smoothing spline \citep{1978pgts.book.....D} to smooth the whole spectrum and then exclude those pixels deviates from the smoothed spectrum by a distance larger than a threshold, e.g., 1.5 times the standard deviation of the residuals in the wavelength bin.
The pseudo-continuum is then estimated by smoothing the reserved pixels in the spectrum.
The softness of the smoothing spline, the distance threshold and the width of the wavelength bins can be adjusted using experience and also according to the properties of the spectral data in hand.
}
Then all stellar spectra are re-sampled to the same wavelength grid.
Assuming that we have $m$ stellar spectra in the training set and each spectrum has $n$ pixels, let $F_{i,j}$ be the $j$th pixel of the $i$th normalized stellar spectrum in the training set, then we have
\begin{equation}
\mu_i={1\over{m}}\sum_{i=1}^{m} F_{i,j}
\end{equation}
and 
\begin{equation}
s_i=\sqrt{{1\over{m-1}}\sum_{i=1}^{m} (F_{i,j}-\mu_i)^2}.
\end{equation}
$F_{i,j}$ is then standardized via 
\begin{equation}
f_{i,j}={{F_{i,j}-\mu_i}\over{s_i}}
\end{equation}
Stellar labels are also standardized in the same way. When the stellar labels are estimated for the observed spectrum in the prediction process, they will be re-scaled back to physical units.

\bz{
It is noted that bad pixels are quite common in spectroscopic surveys due to sky subtraction, cosmic rays  and problems occur in data reduction.
These bad pixels contain no information about stars and their errors can not be estimated, so that they should be excluded in our analysis.
Usually they flagged in the released spectral data products by setting flux error to a very large number or assigning a special flag.
In particular, in our analysis, the LAMOST spectra provide {\ttfamily OR\_MASK} for every pixel in a spectrum, which equals to zero when no problems occur in any single exposure and otherwise equals to a non-zero integer depending on the kind of problem it suffers from (cf. \url{http://dr5.lamost.org/doc/data-production-description} for more information).
We exclude those bad pixel with non-zero {\ttfamily OR\_MASK} values by assigning zero weights in the final spectral fitting.
%For good pixels, the weights are set to 1.
}

\subsection{Training}
\bz{
Support Vector Regression (SVR) is a robust non-linear regression method and has been used in many astronomical studies \citep{Liu2012, Liu2015},
particularly in spectral data analysis \citep{2014RAA....14..423L, 2014ApJ...790..105L, 2015MNRAS.447..256B, 2015MNRAS.452.1394L}.
}.
A more complete description of SVR can be found in \cite{ss2004}.
Since \slam\ is implemented with {\ttfamily{python}}, we adopt the python wrapper of LIBSVM\footnote{A multi-programming language package to solve the support vector machine problems, including SVR regression provided by \cite{Chang2011}.} in the {\ttfamily{scikit-learn}} \citep{Pedregosa2012} package for convenience.

There are two free hyper-parameters, $C$ and $\epsilon$, which represent for the penalty level and tube radius, respectively, in the SVR algorithm. Then we adopt the radial basis function as the kernel (RBF kernel, $K(\mathbf{x}, \mathbf{x}')=\exp{\left( -\gamma||\mathbf{x}-\mathbf{x}'||\right) }$) in SVR.
As a consequence, an additional hyper-parameter $\gamma$, which indicates the width of the RBF kernel, also needs to be determined.

The choice of the hyper-parameters, $C$, $\epsilon$ and $\gamma$, sets the complexity of the SVR model.
For example, a large $C$ penalizes outliers heavily so that the regression will probably be very curved to pass through as many data points as possible, while a small $C$ tells SVR to ignore the outliers and follow a smooth trend of the data. %In other words, high (low) model complexities is analogy to high (low) order polynomial.
In \slam, the best values of the hyper-parameters are not freely controlled, but are automatically determined by the training set itself.
In other words, it is the training set itself, not the user, that determines the adopted model (SVR) complexity pixel-by-pixel.

\subsubsection{\textit{Adaptive} model complexity}
% The key to achieve good performance of SVR is to find a proper set of hyper-parameters. Therefore, we implement a grid search to optimize the choice of hyper-parameters. The "optimized" hyper-parameters should be the ones that minimizes the $k$-fold cross-validated mean squared error (CV MSE). In this sense, \slam\ chooses the best-fit hyper-parameters at each wavelength. Because the hyper-parameters are chosen based on the specific training data in the above process, \slam\ indeed has an \textit{adaptive} model complexity for each spectral flux. Therefore, the \textit{over-fitting} and \textit{under-fitting} issues are automatically avoided in \slam\ essentially. 

\bz{
As described above, the $j$th pixel in the training set has a mean of 0 and a variance of 1.
Let $\boldsymbol{\theta}_i$ denote the stellar label vector of the $i$th star in the training set (i.e., a vector consisting of \teff, \logg\, and elemental abundances) and $f_j(\boldsymbol{\theta}_i)$ be the $j$th pixel of the model output spectrum corresponding to the input stellar label vector $\boldsymbol{\theta}_i$.
Once the model is trained via a specific set of hyper-parameters, we are able to evaluate the Mean Squared Error (MSE) and Median Deviation (MD) separately defined as
\begin{equation}\label{eq:mse}
MSE_j = {1\over{m}}\sum_{i=1}^{m} [f_j(\boldsymbol{\theta}_i)-f_{i,j}]^2
\end{equation}
and 
\begin{equation} \label{eq:md}
MD_j={1\over{m}}\sum_{i=1}^{m}[ f_{j}(\boldsymbol{\theta}_i)-f_{i, j}].
\end{equation}
MSE quantifies the scatter of the regression model and MD quantifies the bias.
For the worst regression model, i.e., a constant model, ${\rm MSE}=1$ because it turns out to be the variance of $f_{i,j}$ according to Eq.~(\ref{eq:mse}).
"Theoretically", the smaller MSE is, the better the fitting is.
However, if we train SVR models directly on the whole training set and pursue the model that minimizes both MSE$_j$ and MD$_j$, we probably get an \textit{over-fitted} model which gives us ${\rm MD}_j={\rm MSE}_j=0$ by interpolating data.
In practice, to avoid the \textit{over-fitting} problem, we use the $k$-fold cross-validated MSE (CV MSE) and $k$-fold cross-validated MD (CV MD) instead, i.e., evaluate the Eq.~(\ref{eq:mse}) and (\ref{eq:md}) via though the $k$-fold cross-validation technique.
In this process, the training set is randomly split out into $k$ subset (usually 5 to 10), and the $f_{j}(\boldsymbol{\theta}_i)$ is predicted by an SVR model trained on the other $k-1$ subsets of the training set.
After looping over all subsets, we calculate the MSE$_j$ and MD$_j$ based on these predicted fluxes in cross-validation and the true fluxes in the training set.
To distinguish them from normal MSE$_j$ and MD$_j$ without cross-validation, we name them CV MSE$_j$ and CV MD$_j$, respectively.
}

\bz{
Because \textit{over-fitting} is generally avoided through such a cross-validation technique, we are able to use the CV MSE$_j$ and CV MD$_j$ to reasonably assess how well the SVR with such model complexity can reproduce the spectral flux of the $j$th pixel in the training set.
In particular, CV MD$_j$ is usually very small once a proper model complexity is adopted.
Therefore, the best model complexity (the best set of hyper-parameters) can be determined by searching for the lowest CV MSE$_j$ after looping over all sets of hyper-parameters specified.
Finally, we train SVR with the chosen \textit{best} set of hyper-parameters on the whole training set for this pixel. 
The MSE$_j$ and MD$_j$ of this final model are calculated directly based on the whole training set to quantify the scatter and bias of the SVR with the best model complexity.
The final MSE$_j$ is also adopted as the model error in the later processes.
By doing so pixel-by-pixel, we guarantee the best model complexity for each pixel.
As a comparison, the final MSE and MD of SLAM are compared to \cannon\ in Section \ref{sec:test} to show the improvements introduced by this adaptive model complexity.
And in the Appendix~\ref{app:a} we use mock data to show how to choose the best hyper-parameters from a grid more comprehensively.
}

\subsection{Prediction}
\bz{
With the Bayesian formula, the posterior probability density function of stellar labels given an observed spectrum is shown as the following,
\begin{equation}
p\left(\boldsymbol{\theta} \vert \boldsymbol{f}_{\rm obs}\right) \propto p\left(\boldsymbol{\theta} \right) 
\prod_{j=1}^{n} p(f_{j, \rm obs} \vert \boldsymbol{\theta} ) 
, \\
\label{eq1}
\end{equation}
where $\boldsymbol{\theta}$ is the stellar label vector, $\boldsymbol{f}_{\rm obs}$ is the observed spectrum vector, $f_{j,\rm obs}$ is the $j$th pixel of the normalized observed spectral flux, $p(f_{j, \rm obs} \vert \boldsymbol{\theta} )$ is the likelihood of the spectral flux $f_{j,\rm obs}$ given $\boldsymbol{\theta}$, and $p(\boldsymbol{\theta})$ is the prior of $\boldsymbol{\theta}$.
The estimation of stellar labels can be easily done by maximizing the posterior probability $p\left(\boldsymbol{\theta} \vert \boldsymbol{f}_{\rm obs}\right)$.
Although it is important to set a proper prior of stellar parameters from external source (e.g., the Galactic model, parallax, proper motions), we adopt an uniform prior in this paper for simplicity in \slam. A prior can be added depending on the specific scientific scenario in future works.
}

Adopting a Gaussian likelihood, the logarithmic form of Eq.~(\ref{eq1}) becomes
\bz{
\begin{equation}\label{eq:logpost}
\ln{p\left(\boldsymbol{\theta} \vert \boldsymbol{f}_{\rm obs}\right)} = \\
-\frac{1}{2} \\
\sum_{j=1}^{n} \left\{ 
\frac{\left[f_{j,\rm obs}-f_{j}(\boldsymbol{\theta})\right] ^2}{\sigma_{j,\rm obs}^2+\sigma_{j}(\boldsymbol{\theta})^2} \\
+ \ln{[2\pi (\sigma_{j,\rm obs}^2+\sigma_{j}(\boldsymbol{\theta})^2)]}
\right\}
\end{equation}
}
where $f_{j,\rm obs}$ is the $j$th pixel of the observed spectrum, $f_{j}\left(\boldsymbol{\theta}\right)$ is the output spectral flux given the stellar label vector $\boldsymbol{\theta}$, $\sigma_{j,\rm obs}$ is the uncertainty of the $j$th pixel of the observed spectrum, and $\sigma_{j}(\boldsymbol{\theta})$ is the uncertainty of the $j$th pixel of the output spectrum corresponding to the stellar labels $\boldsymbol{\theta}$.
% Since $\ln{p\left(\boldsymbol{\theta} \vert f_{\rm obs}\right)}$ can be specified up to a constant, given a spectrum and its error, the last term in this equation can be omitted since it's a constant.
% Note that a generated model spectrum has its error $\sigma_{\rm SLAM}$.
% Usually it can be ignored if the observation error $\sigma_{\rm obs}$ dominates.
In practice, $\sigma_{j}(\boldsymbol{\theta})$ is roughly replaced with CV MSE$_j$, which is independent of $\boldsymbol{\theta}$.

\bz{
Using the Markov chain Monte Carlo (MCMC) technique to sample the posterior function Eq.~(\ref{eq:logpost}) for millions of spectra is not practical due to the computational cost, especially when the number of dimensions goes large.
Therefore, we adopt the Maxmum Likelihood Estimation (MLE) method with Levenberg-Marquardt \citep[LM, ][]{1978LNM...630..105M} least squares optimizer in this work.
The initial guess of $\boldsymbol{\theta}$ is determined by comparing the $\boldsymbol{f}_{\rm obs}$ to the training set and picking the one with maximum likelihood.
The outputs are the stellar labels which maximize the likelihood function and the corresponding covariance matrix. The convergence status is also part of the output, and stars will be marked out if SLAM does not converge within the maximum number of iterations.
}

\subsubsection{Uncertainty}
\bz{
The output covariance matrix of \slam\ is converted from the Hessian matrix produced in {\ttfamily scipy.optimize.least\_squares} method in the SciPy package \citep{jones2001}.
We refer to \cite{1978LNM...630..105M} for how the Hessian matrix is calculated. }
The diagonal elements of the covariance matrix are considered as the \textit{formal errors} for the corresponding stellar labels, hereafter we call them \textit{SLAM errors}.

When making prediction for a data set whose true stellar labels are known, we are able to calculate the cross-validated scatter (\textit{CV scatter}) and cross-validated bias (\textit{CV bias}), which are considered as the standard deviation and mean deviation, respectively. Namely,
\begin{equation}
CV\ bias ={1\over{m}}\sum_{i=1}^{m} (\boldsymbol{\theta}_{i, \rm SLAM}-\boldsymbol{\theta}_i)
\end{equation}
and
\begin{equation}
CV\ scatter = {1\over{m}}\sqrt{\sum_{i=1}^{m} (\boldsymbol{\theta}_{i, \rm SLAM}-\boldsymbol{\theta}_i)^2}.
\end{equation}
Note that the CV scatter/bias are statistics of stellar labels, while the CV MSE/MD described above are statistics of stellar spectra. In principle, a good data-driven method has a very small CV bias and CV scatter. To investigate the precision of a data-driven method, the CV scatter should be used because the CV scatter quantifies the \textit{precision}, while \slam\ error represents for the internal uncertainty of the optimization method.

%___________________________________________________________________________________
%
%               Replication LAMOST DR5 and APOGEE DR15
%___________________________________________________________________________________
%

\begin{figure*}
\plotone{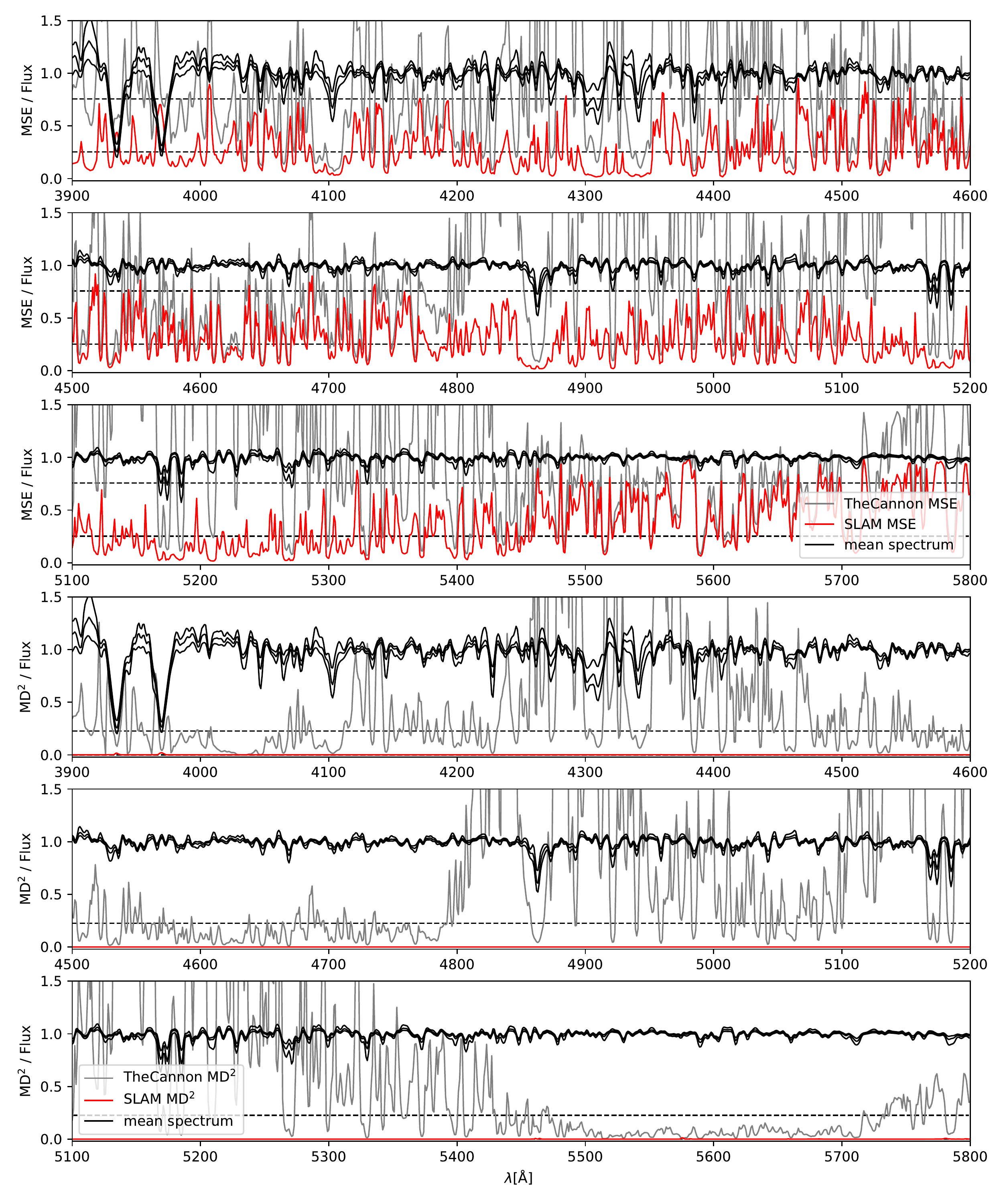}
\caption{This figure shows the comparison of MSE and MD$^2$ of \slam\ and \cannon.
The black lines always represent the 16, 50 and 84 percentile values of each pixel of the spectra of the training set. The red lines in the upper (lower) three panels show the MSE (MD$^2$) from \slam, while the gray lines show similar quantities derived from \cannon. The black dashed lines are the mean level of the MSE/MD$^2$.
\label{fig_mse_comparison}}
\end{figure*}

\begin{figure*}
\plotone{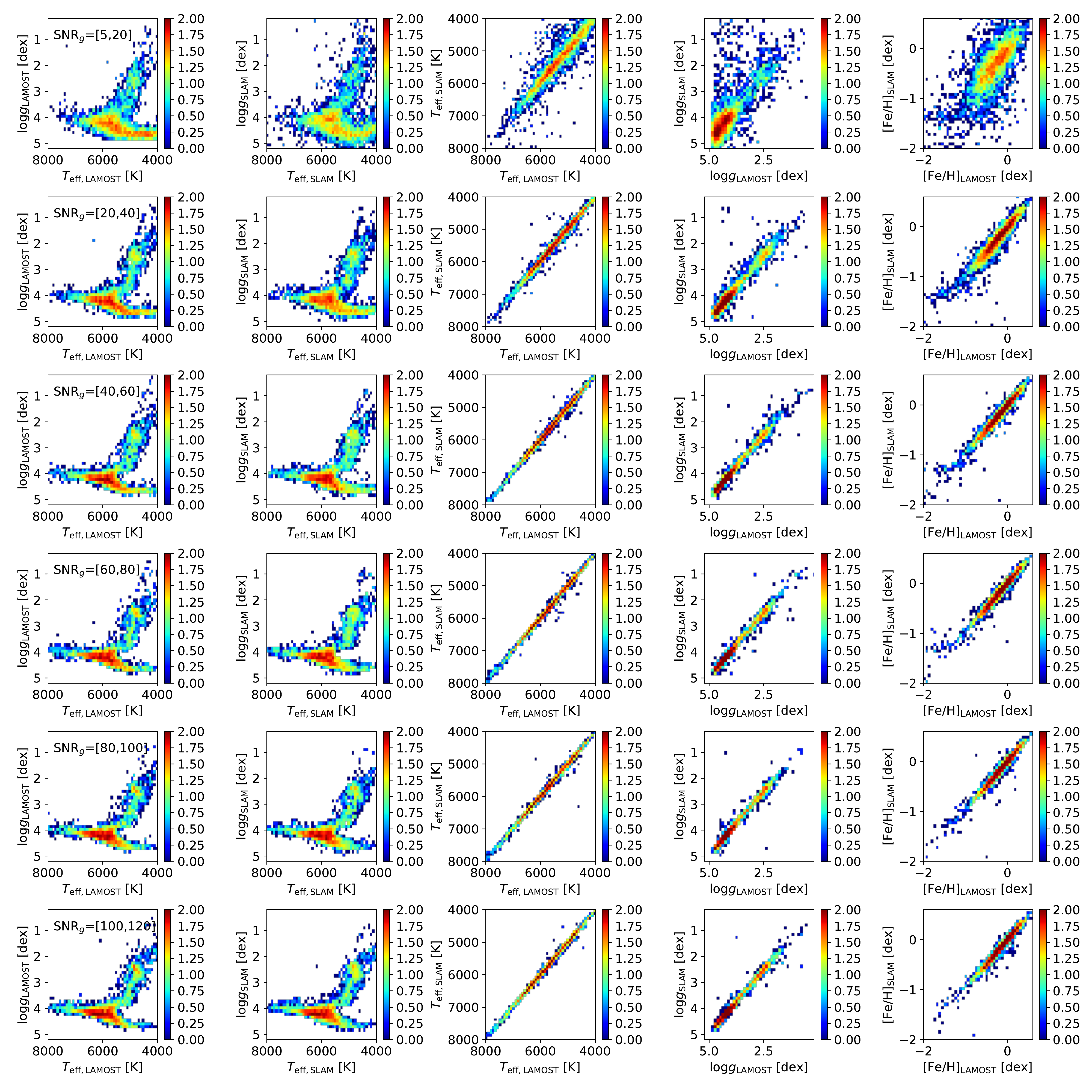}
\caption{This figure shows the distributions of the predicted stellar labels at different ranges of SNR$_g$.
The 6 rows from top to bottom correspond to 6 different SNR$_g$ intervals. In each row, the first panel shows the diagram of \lamost\ DR5 \teff-\logg\ which are regarded as the true values. The second panel shows the similar \teff\ and \logg\ diagram with values derived from \slam. The third, fourth, and the last panels show the \slam-derived stellar labels against the corresponding \lamost\ values. In all panels, color indicated the sample counts in logarithmic scale.
\label{fig_lamost_diag}}
\end{figure*}

\begin{figure*}
\plotone{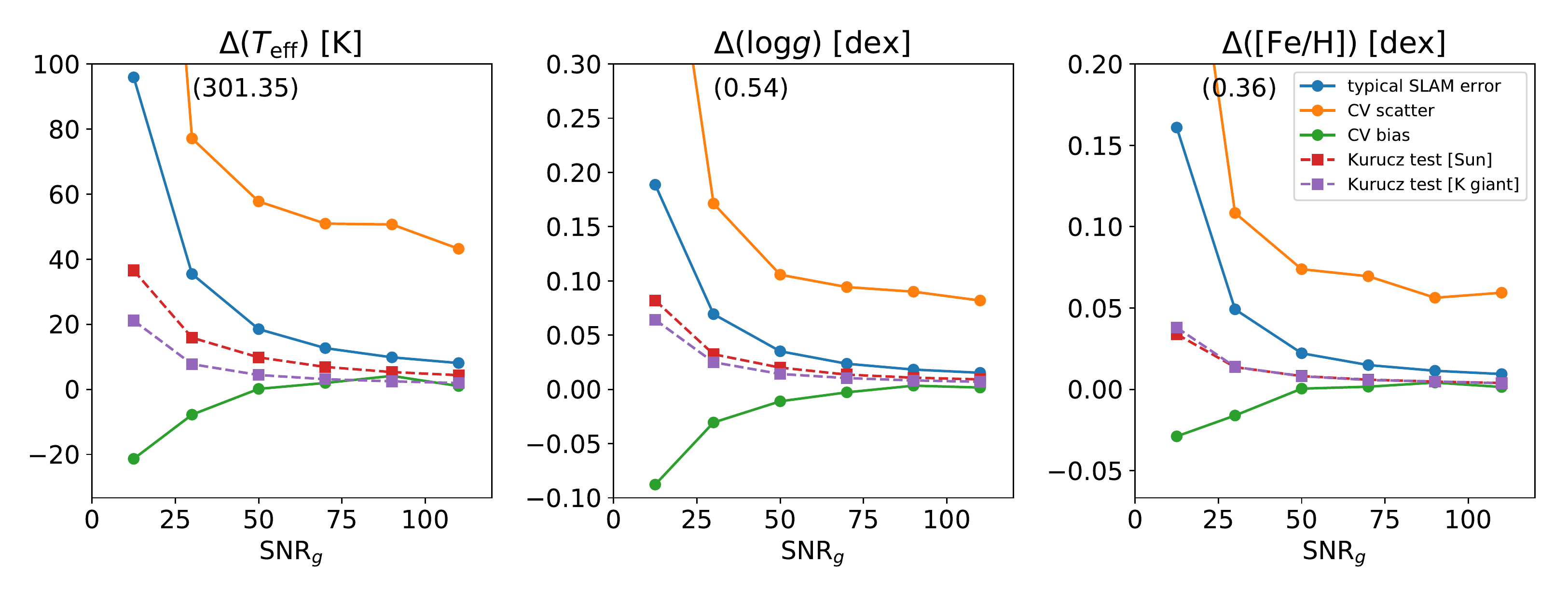}
\caption{This figure shows how the errors of stellar labels change with SNR$_g$.
In all panels, the blue curves represent the \slam\ errors (formal errors). The red and purple curves represent the formal errors of tests for synthetic spectra of solar-like and K giant stars, respectively, selected from the Kurucz ATLAS9 model. The orange and green curves represent the CV scatter and bias, respectively. The first orange points at SNR$_g=12.5$ (corresponding to the $5<{\rm SNR}_g<20$ bin) is located beyond the figure, thus we mark the values of them in brackets at the tops of the panels. Clearly all of them decrease as SNR$_g$ increases. At SNR$_g>100$, the typical CV scatters of \teff, \logg\ and \feh\ are about 50K, 0.10 dex and 0.07 dex, respectively.
\label{fig_lamost_snr}}
\end{figure*}

\section{Tests on LAMOST DR5}\label{sec:test}
\subsection{The LAMOST survey}
The Large Sky Area Multi-Object Fiber Spectroscopic Telescope (LAMOST) telescope, also called the Guo Shou Jing Telescope, is a 4-meter reflecting Schmidt telescope with a 5-degree field of view, on which 4000 fibers are installed. The spectral resolution is R$\sim$1800 covering all optical wavelengths \citep{2012RAA....12.1197C, 2012RAA....12..735D, 2014IAUS..298..310L, 2015RAA....15.1089L}.
%The targets of the LAMOST survey are selected from several catalogs, including UCAC4 \citep{2013AJ....145...44Z}, Pan-STARRS 1 \citep{2012ApJ...750...99T}, the Xuyi Schmidt Telescope Photometric Survey of the Galactic Anti-center (XSTPS-GAC) \citep{2015MNRAS.448..855Y}, and 2MASS \citep{2006AJ....131.1163S}.
\bz{The $r$-band apparent magnitude of the survey covers from 9 to 17.8 mag. In this work, we use the LAMOST Data Release 5 (DR5), which includes observations from September in 2011 to June in 2017.}
The LAMOST DR5 provides $\sim$9 million spectra among which $\sim$5 million are with stellar parameters estimated by LASP \citep{2011RAA....11..924W, 2014IAUS..306..340W}. We use this data set to investigate the performance of \slam\ on dealing with a wide range of \teff\ and compare it with \cannon.

\subsection{Training}
\bz{
The stellar labels of LAMOST AFGK stars used here are estimated with the LAMOST Stellar Parameter pipeline \citep[LASP, ][]{2011RAA....11..924W, 2014IAUS..306..340W} and can be downloaded from \url{http://dr5.lamost.org/}.
}
We firstly select the samples with reliable stellar parameters using the following empirical criteria:
\begin{enumerate}
\item $4000<T_{\rm eff}/{\rm K}<8000$,
\item $0.5<\log{g}/{\rm dex}<5.5$,
\item $-2.5< {\rm [Fe/H]}/{\rm dex}<1.0$,
\item $\Delta(T_{\rm eff})/{\rm K}<200$,
\item $\Delta(\log{g})/{\rm dex}<0.1$,
\item $\Delta({\rm [Fe/H]})/{\rm dex} < 0.1$.
\end{enumerate}
%Empirically LAMOST stellar labels of AFGK stars are acceptable in this range.
Then, we randomly select 5000 training stars with high $g$-band signal-to-noise ratio ($120<{\rm SNR}_g<140$) among them, where SNR$_g$ refers to the average signal-to-noise ratio of the LAMOST spectrum at $g$-band. %These criteria ensure that these spectra have good quality and the stellar labels are reliable.
To validate the model at different signal-to-noise ratios, we separate SNR$_g$ into 6 intervals, i.e. (5, 20), (20, 40), (40, 60), (60, 80), (80, 100), and (100, 120) and randomly select 5000 stars in each of the SNR$_g$ intervals as the test sets.

\bz{All spectra are shifted to the rest frame using the LAMOST DR5 radial velocity and re-sampled to 1.0 $\rm \AA$ step wavelength grid from 3900 to 5800 $\rm \AA$.
And those with more than 50 bad pixels are excluded. The reason of this exclusion is that for these spectra we are not certain about whether their pseudo-continuum estimated is consistent with other spectra or not.} In the training process, we set the grid of hyper-parameters to be $\epsilon=0.05$, $C=$ [10, 100] and $\gamma=$[0.1, 0.01]. Each pixel is trained with SVR and set with the hyper-parameters based on a 5-fold cross validation.
\bz{
$\epsilon$, although as one of the three hyper-parameters, is simply fixed due to the fact that it represents the tube radius of the SVR outside of which the SVR regard data as outliers and ignores them.
The role of $\epsilon$ is very like a tolerance of the regression function.
Taking this training set as example, the typical standard deviation of the normalized flux is about 0.02. Therefore, the tube radius stands for $0.02 \times 0.05=0.001$ in normalized flux (recall that the normalized flux is standardized to have a variance of unity and SVR works in the standardized space).
So the typical "S/N" ratio of a spectrum predicted by SVR model is around $1/0.001=1000$.
In other words, the SVR could reproduce spectra at a "S/N" ratio at 1000.
Decreasing $\epsilon$ to an even lower value to raise this “S/N” ratio is not necessary as in our test we typically work at S/N$\sim100$, while increasing $\epsilon$ could make our training more computationally expensive and easily get \textit{over-fitted}.
For the $C$ and $\gamma$, it could be inferred that setting them too large or too small is unnecessary in the standardized space from the Appendix.
As a test, we try to make the grid of $C$ and $\gamma$ very sparse and see how good the results could be.
}
%However this sample is unbalanced in parameter space.
%For example, there are more main sequence stars than red giant stars.
%Then we reduce the training set by making the sample uniform in parameter space, i.e., we randomly select only 2 stars in each 3-dimensional bin with widths of 100 K in \teff, 0.1 dex in \logg\ and 0.1 dex in \feh.
%At last only $\sim$8000 stars are used in training.

We also train \cannon\ with the same training set for comparison and plot both the training MSE of \slam\ and \cannon\ in Figure~\ref{fig_mse_comparison}. The black lines shows the 16, 50 and 84 percentile of the training spectra. The median one represents the 'typical' spectrum of the training sample. In the upper three panels, the red and gray lines show the CV MSE of \slam\ and \cannon, respectively, while in the lower three panels the red and gray lines denote the CV MD squared of \slam\ and \cannon, respectively. The two black dashed lines represent the median values of the red and gray lines.
\bz{
We found that \slam\ is able to make the CV MSE much lower than that of \cannon\ at almost all wavelengths.
The reason is that the quadratic model adopted by \cannon\ is insufficiently flexible to model spectra in such wide ranges of stellar labels like $4000<T_{\rm eff}/{\rm K}<8000$. And the CV MD$^2$ of \slam\ also show much improvements compared to that of \cannon.
}

\subsection{Prediction}
In the first row of Figure~\ref{fig_lamost_diag}, we show the \teff-\logg\ distribution of the training sample stars with $5<{\rm SNR}_g<20$ in the first panel and the \slam-predicted \teff\ and \logg\ in the second panel.
In third, fourth and fifth panel, we show the diagonal plot of the \teff, \logg\ and \feh, respectively, to compare the estimates from \slam\ with the originals of LAMOST. From the second to the last row, we show similar plots for stars with $20<{\rm SNR}_g<40$, $40<{\rm SNR}_g<60$, $60<{\rm SNR}_g<80$, $80<{\rm SNR}_g<100$ and $100<{\rm SNR}_g<120$, respectively.
As ${\rm SNR}$ increases, the \slam-predicted values become more and more consistent with the true values.

In Figure~\ref{fig_lamost_snr}, we show the SLAM errors, CV scatter, and CV bias at various SNR$_g$.
Note that the SLAM errors are very small compared to CV scatter. For stars with SNR$_g>100$, the SLAM errors for \teff, \logg\ and \feh\ are smaller than 10 K, 0.03 dex, and 0.02 dex, respectively. We also show the simulated error values for a solar-like star and a K giant star at different SNR$_g$ using the ATLAS9 synthetic spectra \citep{2003IAUS..210P.A20C}. Although the SLAM errors of the observed spectra are very small, they are much larger than the simulated values, which can be regarded as the lower limits of errors.

On the other hand, the CV scatters are larger than the SLAM errors. At high SNR$_g$ end, the CV scatters of \teff, \logg\ and \feh\ are $\sim$50K, 0.10 dex, and 0.07 dex, respectively. These values are very similar to the values reported in \cite{Ho2017b}. However, it is worth to note that our sample is distributed in wider ranges than that studied by \cite{Ho2017b}, which only considered red giants with low effective temperatures. In general, the hot and warm stars may suffer from larger uncertainties of stellar parameter estimates than the cool stars \citep{Liu2012}.

The reason that the SLAM errors are substantially smaller than the CV scatters is because we assumed that both the spectral fluxes and the stellar labels in the training set are infinitely accurate. When we model fluxes as functions of stellar labels, the observed fluxes of the training stars are composed of noise, i.e. \bz{$\boldsymbol{f}_{\rm obs}=\boldsymbol{f}(\boldsymbol{\theta})+\epsilon$},  where $\epsilon$ denotes noise. Meanwhile, the errors in stellar labels are also not taken into account in the model.
%So that the errors of stellar labels will induce some uncertainties in our model.
%While fitting a spectrum is very precise, the uncertainty in our model still exists.
If we train our model with different training samples, the predicted stellar labels would be different due to the different errors implied in the training set. This difference should be larger than the \slam\ error which is internal. 

Another reason is that the errors of stellar labels in the validation sample also exist. This can increase the CV scatter to some extent. For instance, if the stellar labels of the validating sample have errors of 30K in \teff, it is impossible to decrease CV scatter to under 30K.

Therefore, to assess the performance of a data-drive method, CV scatter is the fair quantity rather than the \slam\ error (or the internal error of the method), since the former has taken into account the uncertainties contributed by the training set.%  However, to investigate the precision of a data-driven method, the CV scatter should be used because the CV scatter quantifies the \textit{precision}.

\bz{
\subsection{A comparison to the LASP}
\slam\ is different from LASP \citep{2011RAA....11..924W, 2014IAUS..306..340W} in many different aspects. Since LASP, which uses Ulyss \citep{2009A&A...501.1269K} to predict stellar labels, builds a polynomial model of spectral flux on ELODIE spectral library \citep{2007astro.ph..3658P}, SLAM offers several advantages over it.
The first is that SLAM offers more flexibility and adaptive model complexity taking advantages of the RBF kernel. Secondly, we made SLAM open source and users can choose whatever they want as training set rather than sticking to ELODIE. The third is that SLAM can generally provide uncertainty estimates of stellar labels by applying the relation between the CV scatters and S/N ratios. Last, but the most important difference, is that SLAM is able to extend to more stellar labels, e.g., [$\alpha$/Fe] and other element abundances, while it is impossible with LASP currently.
}

%___________________________________________________________________________________
%
%               APOGEE DR15
%___________________________________________________________________________________
%

\section{Predict stellar labels for LAMOST spectra based on APOGEE DR15}
\label{sec:apogee}
\subsection{The APOGEE survey}
The APOGEE survey provides high-resolution (R$\sim$22,500) $H$-band (15200-16800 \AA) spectra \citep{2017AJ....154...94M}. APOGEE DR15 comprises $>$270,000 high signal-to-noise ratio spectra.
Its pipeline, ASPCAP \citep{2016AJ....151..144G}, produces estimates of the basic stellar labels, abundances, and micro-turbulence. In this section, we use the APOGEE DR15 \citep{ 2018AJ....156..125H} stellar labels in the training set to set up the \slam\ model and predict stellar labels for the LAMOST DR5 low resolution spectra.

\subsection{Training and test set}
We first select our training set from the 86,552 common stars between APOGEE DR15 and LAMOST DR5 by adopting the following criteria,
\begin{enumerate}
\item the signal-to-noise ratio of the APOGEE spectra ${\rm SNR_{APOGEE}} >100$,
\item ${\rm SNR}_{g} >40$ for LAMOST spectra,
\item the ASPCAP stellar label flag {\ttfamily ASPCAPFLAG}$=0$,
\item the ASPCAP effective temperature $3000<T_{\rm eff,APOGEE}/{\rm K}<5500$,
\item the ASPCAP surface gravity $-1<\log{g}_{\rm APOGEE}/{\rm dex}<5$,
\item the ASPCAP overall metallicity $-4.0<{\rm [M/H]/{\rm dex}}<2.0$,
\item the ASPCAP carbon abundance $-0.4<{\rm [C/Fe]}/{\rm dex}<1.0$,
\item the ASPCAP nitrogen abundance $-0.5<{\rm [N/Fe]}/{\rm dex}<1.0$,
\item and the difference of the corresponding LAMOST and APOGEE effective temperature $|T_{\rm eff,APOGEE}-T_{\rm eff,LAMOST}|/{\rm K}<800$.
\end{enumerate}
The purpose of the last criterion is to give a loose condition on the consistency between the stellar labels provided by LAMOST and APOGEE, so that the selected stars have reliable stellar label values. With these criteria, we obtain 17,703 common stars with reliable stellar labels.
\bz{
Then we exclude the LAMOST spectra containing more than 50 bad pixels} and obtain 17,623 stars. 
The grid of hyper-parameters $C$ and $\gamma$ are set to be uniform in logarithmic scale, i.e., $C=10^{[0. , 0.5, 1. , 1.5, 2. ]}$ and $\gamma=10^{[-3. , -2.5, -2. , -1.5, -1. ]}$, while $\epsilon$ is fixed at 0.05.
\bz{
We use a leave-$\frac{1}{10}$-out training process to exclude the stars whose stellar labels deviate from the training set values by more than 4 times the standard deviation stellar label residuals in any dimension. Finally, our training sample contains 17,175 stars and then we train SLAM on this data set.}
%The hyper-parameters $C$, $\epsilon$ and $\gamma$ are chosen to be 2.0, 0.02 and 2.0 respectively.
%As the number of dimensions is larger, we didn't show the single-pixel diagnostic diagram for difficulties in visualization.
%Instead we show the fitting results for a randomly chosen star, 2M05023229+2335361, as an example.
%For the upper sub-figure, the first and second panels are mean and scale spectrum derived from training sample, respectively.
%The typical scale here is 0.024.
%The third panel shows the fitting result while the fourth panel is the corresponding residual spectrum.
%The fitting result is good since the standard deviation of residual spectrum is only 0.007, which is quite close to the ideal tolerance $\epsilon \times \rm{typical~scale} \approx 0.005$, indicating the training process is quite good and prediction is quite successful considering that the SNR=237 also introduces flux error $\sim$ 0.005.
We use an 8-fold cross-validation to find the best-fit hyper-parameters and conduct the training process. 
Then we apply the tuned \slam\ model to all 8,171,443 stars ({\ttfamily class=STAR} in LAMOST catalog) in LAMOST DR5. \slam\ successfully converges for 5,132,474 stars.

\begin{figure*}
\plotone{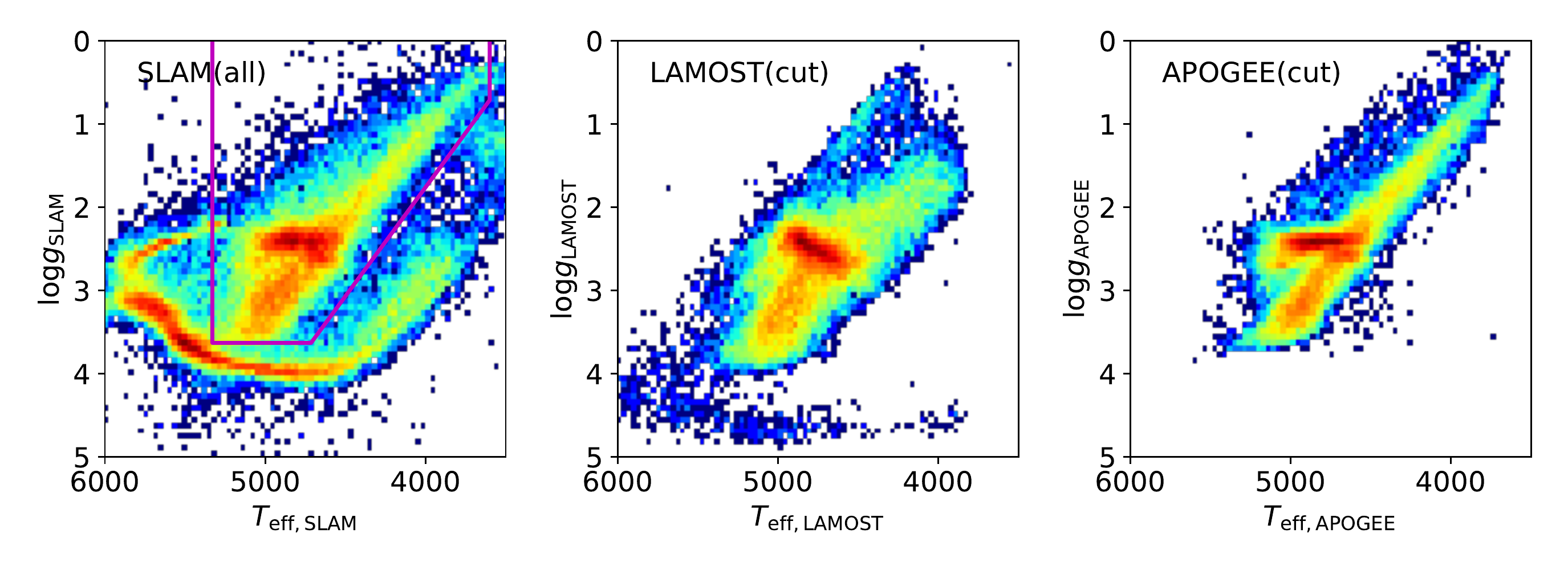}
\caption{The left panel shows the distribution of \slam-predicted \teff\ and \logg\ of all converged LAMOST DR5 stars. The pink solid polygon represents the selected area for K giant stars.
The middle/right panel shows LAMOST/APOGEE \teff-\logg\ diagrams for the sample located in the pink polygon.
\label{fig_laap_giant_branch}}
\end{figure*}

In the LAMOST--APOGEE common samples (86,552), \slam\ converged for 57,703 of them and derived their stellar labels. In the left panel of Figure~\ref{fig_laap_giant_branch}, we show their distribution in \teff--\logg\ plane. The \slam-predicted stellar labels has a red giant branch and a stripe, which looks like a distorted main sequence on which most stars do not have APOGEE stellar labels. The stellar labels of the objects located in the stripe are unreliable because that the stellar labels are too far away from the stellar label range of our training set. We cannot apply the simple parameter cut described in \citet{Liu2014} because the stripe turns upward at $T_{\rm eff}<4500$\,K. Therefore, we empirically set up a polygon (shown in pink in the figure) for the selection.
The sample stars located in the pink solid polygon are selected as the K giant stars with reliable stellar labels. We show the corresponding LAMOST (APOGEE) stellar labels of the samples located in the polygon in the middle (right) panel. 

To assess the completeness of our cut, we select stars with the similar criteria listed in the very beginning of this sub-section but ignore the constraints depending on LAMOST stellar labels. Then we have $\sim$22,000 stars with good APOGEE stellar labels left. We check whether these known K giant stars are selected by the empirical polygon and find that the polygon cut loses $\sim$500 K giants, which gives a completeness of about 97\% for the K giant stars.

\begin{figure*}
\plotone{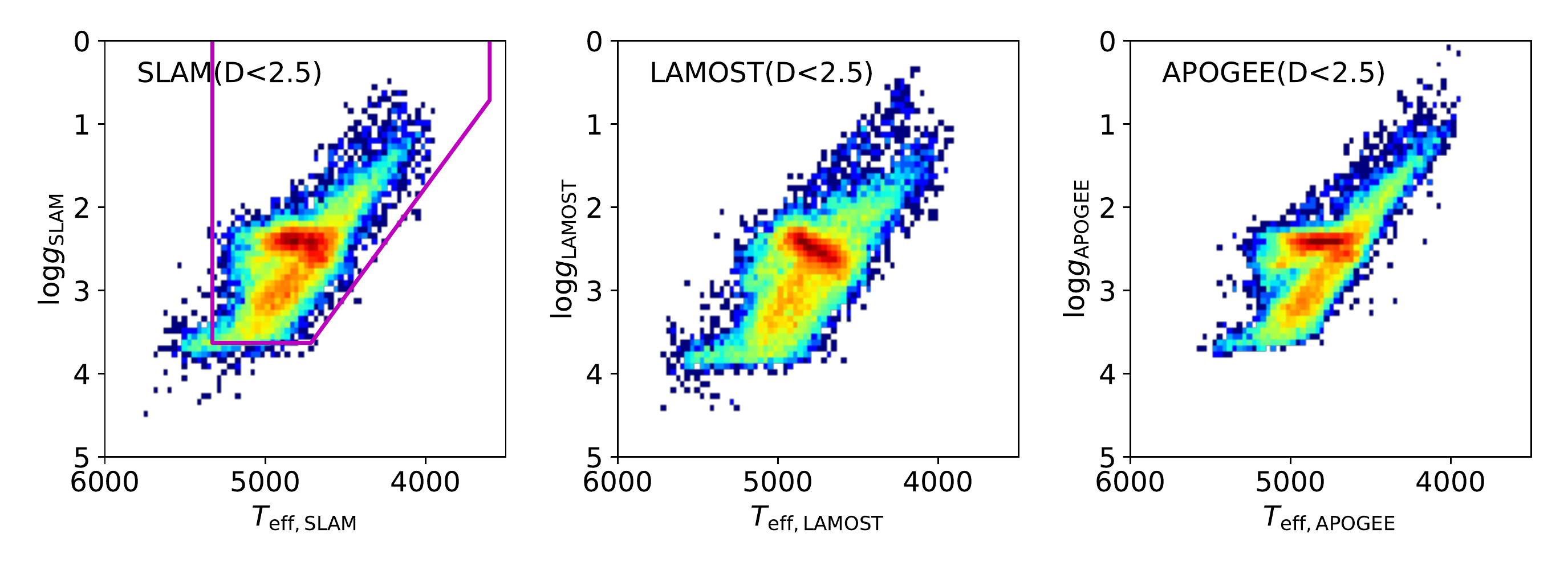}
\caption{The left panel shows the distribution of \slam-predicted \teff\ and \logg\ of LAMOST DR5 stars with $D<2.5$. The pink solid polygon is the same as Figure \ref{fig_laap_giant_branch}.
The middle/right panel shows the LAMOST/APOGEE \teff\ and \logg\ for the same samples.
\label{fig_laap_giant_branch_dho}}
\end{figure*}

We also calculated the "label-distance" $D$ defined by \cite{Ho2017a}, i.e.,
\begin{equation}
\begin{split}
D\ =\ &(T_{\rm eff,SLAM}-T_{\rm eff,LAMOST})^2/(100\ {\rm K})^2 \\
&+(\log{g}_{\rm SLAM}-\log{g}_{\rm LAMOST})^2/(0.2\ {\rm dex})^2 \\
&+({\rm [M/H]}_{\rm SLAM}-{\rm [Fe/H]}_{\rm LAMOST})^2/(0.1\ {\rm dex})^2.
\end{split}
\end{equation}
To be consistent with \cite{Ho2017a}, we consider stars with $D<2.5$ as K giant stars. In the left, middle, and right panels of Figure~\ref{fig_laap_giant_branch_dho}, we show plots similar to Figure~\ref{fig_laap_giant_branch}.
The distribution of the \slam-predicted stellar labels is quite similar with the sample selected using the polygon cut. To select a K giant samples with reliable stellar labels, we suggest one to either using the polygon cut, or using "label-distance" method, or a combination of them.

\subsection{Performance}
\begin{figure*}
\plotone{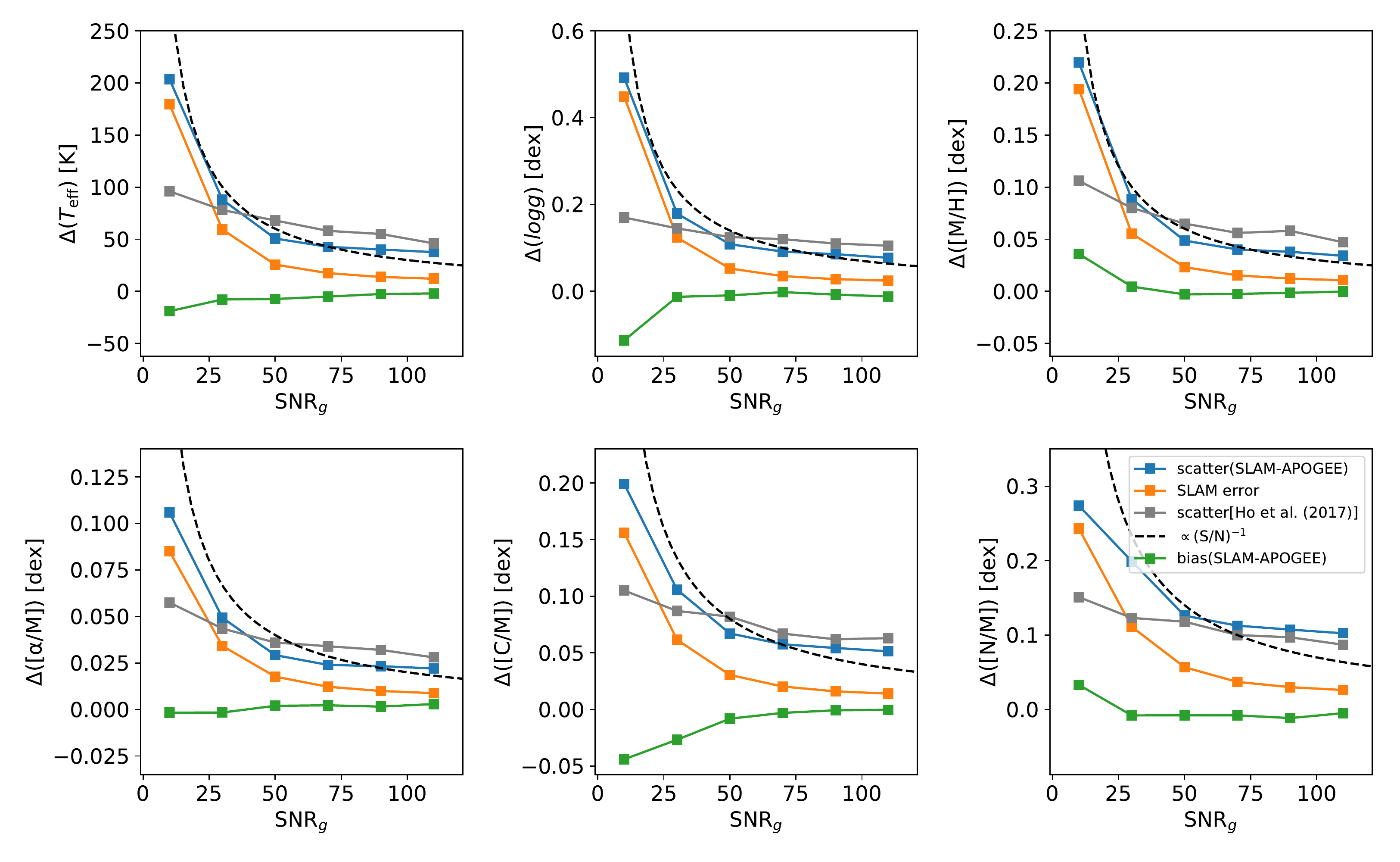}
\caption{This figure shows the comparison of the CV scatters of stellar labels between \slam(blue) and \cannon(gray). \slam\ errors and bias are also shown with green and orange lines, respectively.
The inverse signal-to-noise ratio trends are also superposed with black dashed line.
\label{fig_laap_scatter}}
\end{figure*}

Figure \ref{fig_laap_scatter} shows the CV scatter of the \slam-predicted stellar labels for the LAMOST--APOGEE common stars at different signal-to-noise ratio intervals.
As SNR$_g$ increases, the CV scatters decrease rapidly (shown by blue line) as expected.
At high SNR$_g$ end, the CV scatters of estimated stellar labels are 49 K, 0.10 dex, 0.037 dex, 0.026 dex, 0.058 dex, and 0.106 dex for \teff, \logg, \mh, \am, \cm\ and \nm, respectively. These values are quite similar to those reported in \cite{2017ApJ...843...32T}. Compared to CV scatters, the biases are only as large as one fourth of the scatters at most and thus do not contribute a lot in the total uncertainties. SLAM errors are again much smaller than the corresponding CV scatters. 

\bz{
We found that, although the CV scatters are smaller than that in \cite{Ho2017b} at high SNR$_g$ end, they are much larger at low SNR$_g$ end. 
One probable source of this difference is that in our result, only the 3900 to 5800 $\rm \AA$ part of the LAMOST spectra are used. And we didn't utilize information from photometry.
Our CV scatters are more similar to the inverse of SNR$_g$ trend, which is more realistic for a general test sample. According to the correlations between the CV scatters and SNR$_g$, we suggest that the carbon and nitrogen abundances derived by \slam\ can only be used for stars with SNR$_g>40$.
%This is because that \cite{Ho2017b} removed outliers in their samples when evaluating the scatters, which makes their CV scatters smaller at low signal-to-noise ratio end. 
We also noticed that the bias is significant at the low SNR$_g$ end.
We will try to overcome this problem in our future work.
}

\begin{figure*}
\plotone{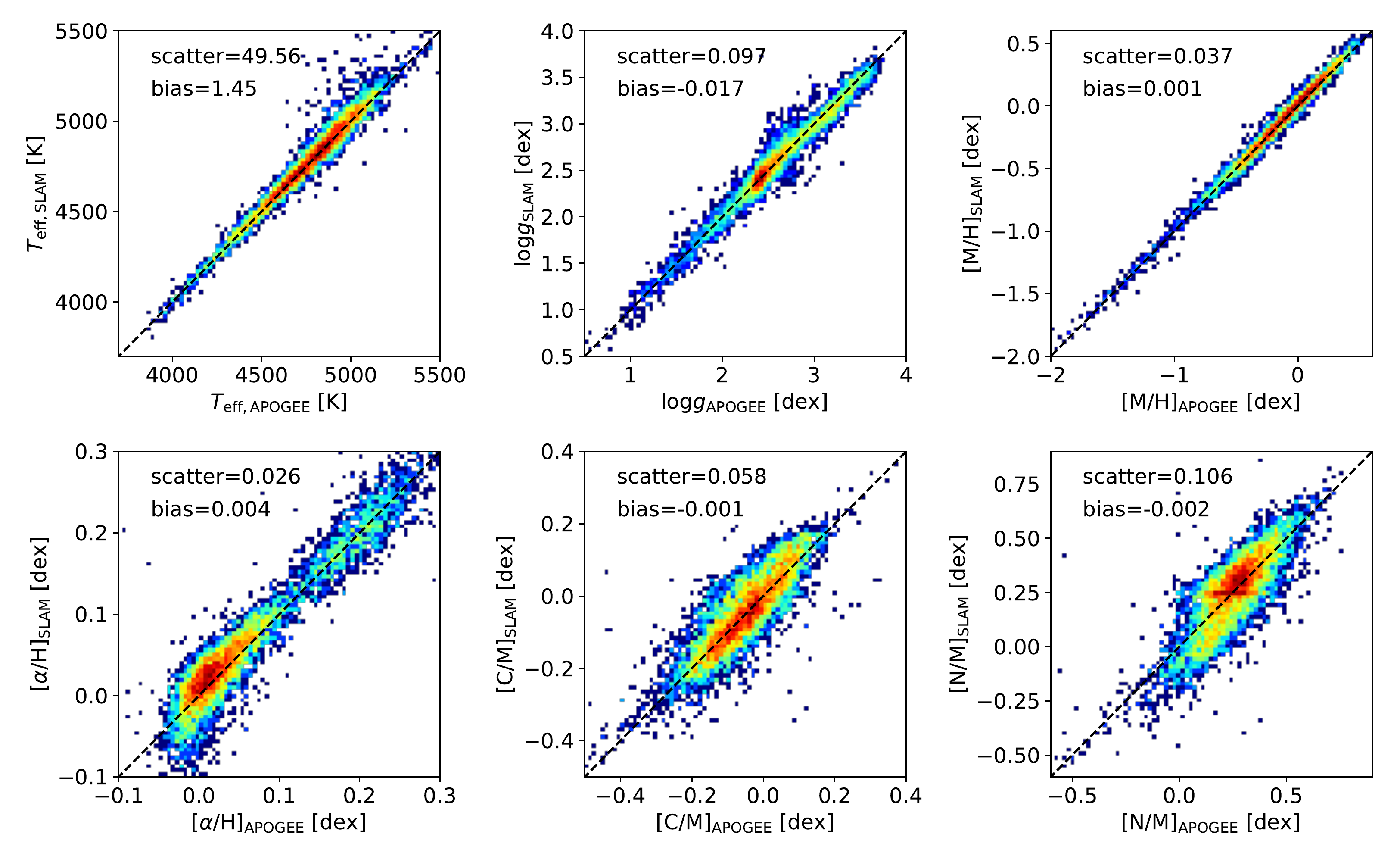}
\caption{This figure shows the diagonal plots of the 6 stellar labels (effective temperature, surface gravity, metallicity, $\alpha$-element abundance, carbon abundance, and nitrogen abundance) for the LAMOST--APOGEE common stars with SNR$_g>100$.
\label{fig_laap_diag}}
\end{figure*}

In Figure \ref{fig_laap_diag}, we show the diagonal plot of our stellar labels against corresponding APOGEE stellar labels for the sub-sample with SNR$_g>100$. It is seen that the \slam-derived \teff, \logg, \mh, \am\, \cm, and \nm\ well agree with the APOGEE values. %Therefore the \slam\ derived stellar labels are valuable data set for the studies of the Milky Way.
%\mh\ down to -2 enables us to study the halo and other substructures of the Milky Way.

\begin{figure*}
\plotone{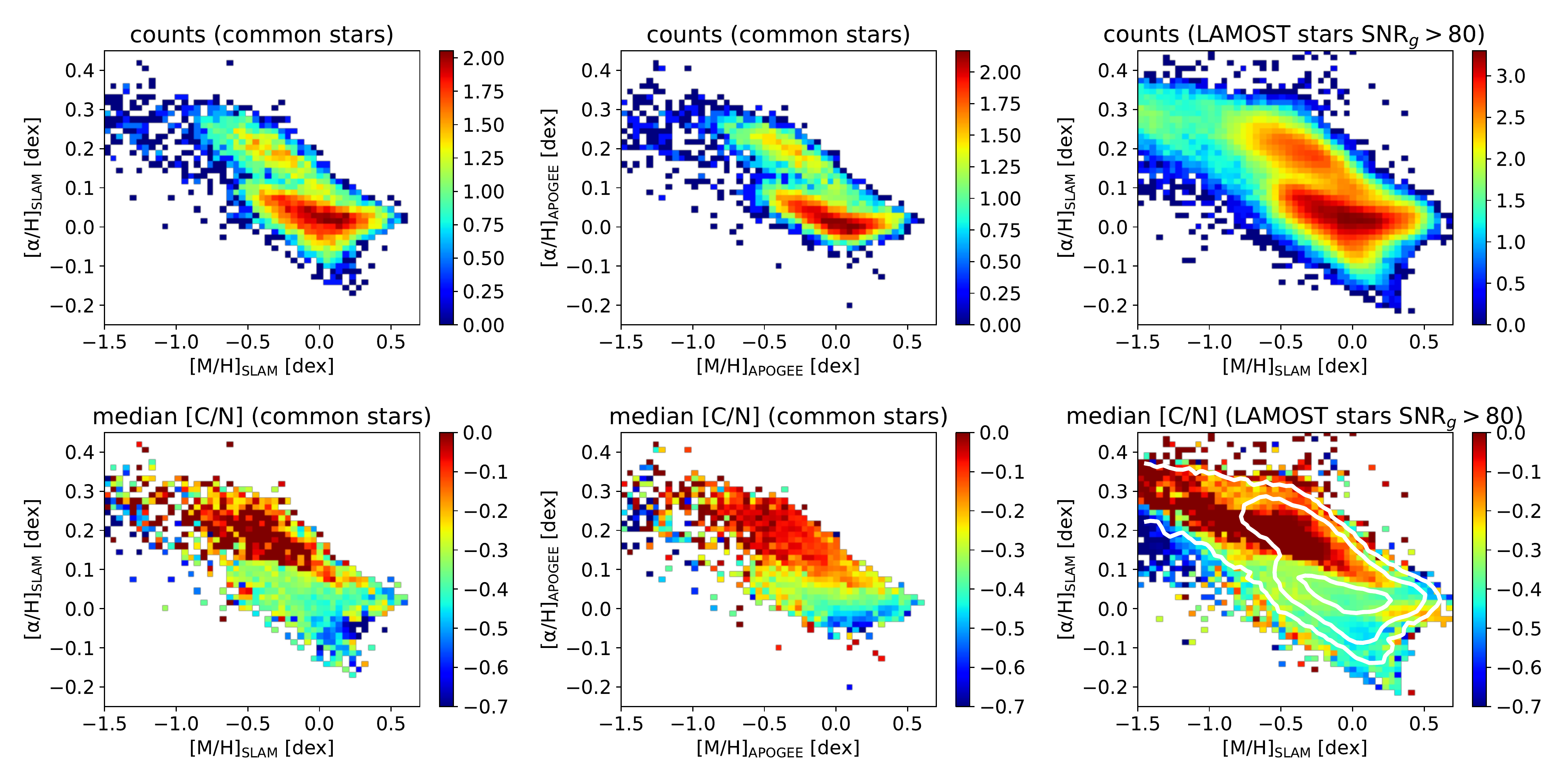}
\caption{The top-left panel shows the distribution of \am-\mh\ plane for the \slam-derived labels for the LAMOST--APOGEE common stars. The top-middle panel shows the similar plot but with APOGEE parameters. The top-right panel shows the similar plot as the top-left panel for all the LAMOST K giant stars with SNR$_g>80$.  The bottom panels show the distributions of median [C/N] in \am-\mh plane. Similar to the top panels, from left to right are the \slam-derived stellar labels for the LAMOST--APOGEE common stars, the APOGEE labels for the common stars, and the \slam-predicted labels for all LAMOST K giant stars with SNR$_g>80$.  In the bottom right panel, white contours of counts are superposed.
\label{fig_laap_macn}}
\end{figure*}

In Figure \ref{fig_laap_macn}, we show the comparison of color coded median [C/N] in the \mh-\am\ plane.
In the left column, the top and bottom panels show logarithmic counts and median [C/N] of \slam\ results, respectively, in the \mh-\am\ plane using the LAMOST-APOGEE common stars. All the stellar labels in the two panels are from \slam. The middle column shows similar plots but using APOGEE stellar labels. In the right column, the two panels show the \slam\ results for all LAMOST K giant stars with SNR$_g>100$. It is seen that the predicted \mh, \am, and [C/N] are similar to the training data.
Meanwhile, [C/N] is in the reasonable range at high density region.

In the final catalog, the output errors of stellar labels are approximated from SNR$_g$ using the empirical function, $a\exp{(-b\times{\rm SNR}_g)}+c$. The best-fit coefficients in the empirical functions are listed in Table \ref{tab:fitting_coef}.
\bz{
The whole catalog will be available with the journal once this article is accepted, and a copy will be hosted at the China-VO Paper Data website \url{http://paperdata.china-vo.org/}. And an example of it is shown in Table \ref{tab:example_table}}.

%Compared to \cannon, as \citep{ness2015} described that the error of \teff, \logg\ and \feh\ using \cannon\ for spectra with highest SNR are 73 K, 0.18 dex and 0.11 dex, respectively, our precision is obviously better because for stars with SNR$\sim$100 \slam's predictions can reach such precision.

\begin{table*}
	\centering
	\caption{The fitting coefficients $a$, $b$, and $c$ for each stellar label. }
	\label{tab:fitting_coef}
\begin{tabular}{ccccccccc}
\hline\hline
stellar label & $a$ & $b$ & $c$ \\
\hline
\teff/K   & 204.8 & 0.056 & 38.8 \\
\logg/dex & 0.592 & 0.063 & 0.069 \\
\mh/dex   & 0.431 & 0.073 & 0.029 \\
\am/dex   & 0.090 & 0.049 & 0.019 \\
\cm/dex   & 0.152 & 0.043 & 0.040 \\
\nm/dex   & 0.152 & 0.031 & 0.072 \\
\hline
\end{tabular}
\end{table*}

%___________________________________________________________________________________
%
%               Discussion
%___________________________________________________________________________________
%

\section{Discussion}
\label{sec:discussion}
%Now we have already demonstrated the capability of \slam\ in deriving reliable stellar labels in a wide range of ranges. However, challenges and issues in applications of \slam\ and other data-driven methods originate from various factors, among which the most important ones should be the pre-processing and training issues.
Although the performance of \slam\ has been well illustrated in sections 3 and 4, \bz{several} challenges and issues, most of which appear in data-driven approaches quite commonly, are worth to be discussed here.

\subsection{Pre-processing}
In the pre-processing step, \slam\ and other data-driven methods operate with RV-corrected and normalized spectra. Consequently, uncertainties in these processes propagate to the final results. However, it is extremely difficult, if it is possible, to automatically determine the proper and consistent pseudo-continuum in the normalization process for various types of stars. Hence, the normalization process induces a certain amount of uncertainties in the normalized spectra, especially for the late type stars (\teff$<4500$ K).
%Therefore, this normalization problem prevents the application in wider range in \teff.

In low resolution spectra, the blending of spectral lines and molecular bands, such as the G band, also increases the uncertainties of the normalized spectra. Weak lines could be overwhelmed by the inconsistency of the normalization. In some cases, the inconsistent normalization may lead to the failure of the stellar label estimation in the data-driven methods.

Although, for K giant stars, the normalization pre-processing in both \slam\ and \cannon\ seem adequate and may not affect the final performance, we should be cautious to this issue, especially when the normalization may induce a variation/deviation larger than the typical training precision.

% Around strong lines, for example, H$\beta$, it is not a problem because the trend is usually clear and a model can easily capture the variations against stellar labels (mainly \teff).
% For weak lines, the variation of flux against stellar labels could be overwhelmed by the inconsistency of pseudo-continuum normalizations and therefore it is hard to estimate weak stellar labels (e.g., nitrogen abundance).
\subsection{Training}
In the training step,  the most important issue is the limited coverage of the parameter space of the training sample. This is also described in section 5.5 in \cite{Ness2015}. Therefore, the selection of training set is crucial.
Once some types of input stars are not included in the training set, the program would not derive meaningful stellar labels.

The second issue is the imbalance of the training sample. Usually very few stars are located near the edge of the parameter space. For example, extremely hot/cool stars and very metal-rich/metal-pool stars are rare. Their spectra are very different from those of normal stars and thus play more important roles in the training process. These stars are anchors which define the edge of the parameter space. However, their small numbers may not effectively leverage the objective function compared to the majority of the normal stars.
% Therefore, in the training step of \slam, adding a proper leverage for these edge stars and other important stars is a possible way to improve the performance.

\bz{
The third issue arises in the flux model. The flux model of \slam\ does not make use of the uncertainties of the stellar labels in the training set.
%can be represented by ${\rm flux}=f(labels)+\epsilon$, where $\epsilon$ denotes random noise. This means that we assume the stellar labels in the training set are accurate and do not have errors, which is not true in reality.
This leads to the underestimation of uncertainties of both the spectra and stellar labels in the model.
To take into account the stellar label errors in the training set, one possible solution is to cross-validate the training samples and get different models using different subsets of the training set, and derive the deviations of predictions using these models.
However, so far it is difficult for us to conduct such a complicated training process due to the high computational expenses.
}

\bz{
\subsection{Computational cost of \slam}
Although SVR is a powerful tool, its computational cost and storage requirements increase rapidly with the number of training vectors \citep{Pedregosa2012}. 
The complexity of the problem solved with LIBSVM scales between $\mathcal{O}(n_{features} \times n_{samples}^{2})$ and $\mathcal{O}(n_{features} \times n_{samples}^{3})$ \citep{Chang2011}, which means that adding more stars in the training set is more difficult than adding more stellar labels.
A fiducial cost of SLAM is that, in our experiment on transferring stellar labels from APOGEE DR15 to LAMOST DR5 in Section~\ref{sec:apogee}, the training takes about 1 day in using an Intel Xeon CPU E5-2690v4 (2.60GHz).
The training cost is also proportional to the number of pixels and the size of the hyper-parameter grid that are tried.
And for prediction, it takes $<$1 min to predict the 6 stellar labels for a spectrum with modest S/N.
For those with very low S/N ratio, it sometime does not converge so that it takes much longer typically.
Therefore, users should be cautious of the computational expense when using SLAM to derive a large number of stellar labels.
}

%\subsection{Training on model spectra}
%\citep{ness2015} proposed that \cannon\ is able to learn from any set of spectra with known stellar labels associated with them.
%However, in the case of model stellar spectra, we need some additional care of the pre-processing step and the training.
%This is because the model stellar spectra are made with no noise.
%As a regression learning model, \cannon\ assumes that random noise is exited in training spectra, so that the trained model can approach the \textit{true} values.
%So one solution is to add reasonable noise on the model spectra before the training step.
%
%Although noise is added, however, we should still guarantee that the training sample is a set of spectra with relatively high \sn\ ratio.
%Therefore, the training step can only accept a very small tolerance.
%In the case of SVR/SLAM, a high C and a low $\epsilon$ is required.

\section{Learning from data: coefficients of dependence (CODs)}
\label{sec:cods}
\begin{figure*}
\plotone{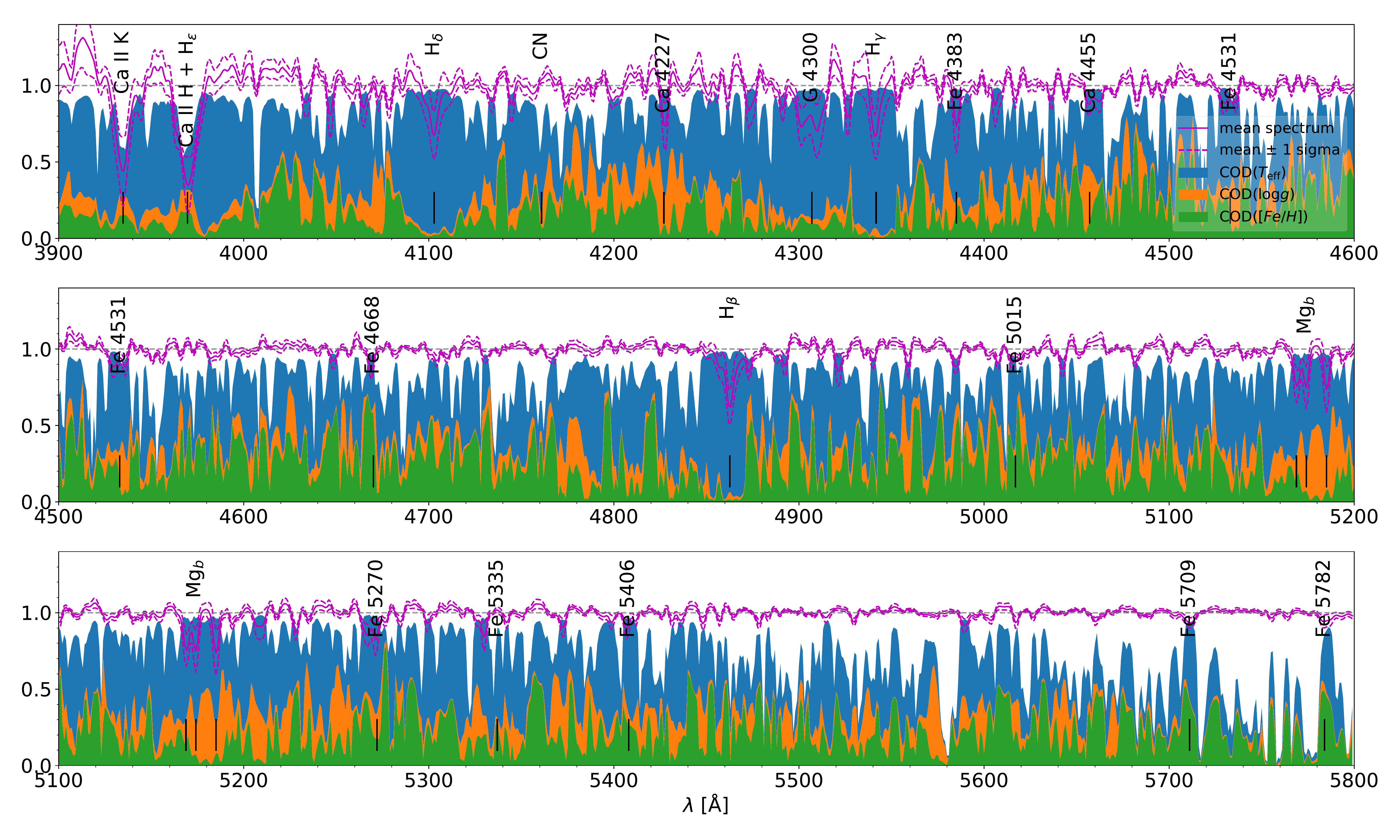}
\caption{CODs from training set with LAMOST spectra and stellar labels.
The pink solid and dashed lines are the 50, 16, and 84 percentiles of normalized spectral fluxes. The blue, orange, and green filled regions represent for the COD(\teff), COD(\logg), and COD(\feh), respectively. %Therefore the upper edge of the blue area is the 1-MSE spectrum.
\label{fig_lamost_cods}}
\end{figure*}

In this section, we present the coefficients of dependency (CODs), which enable us to better understand why machine learning methods generally agree with our experience in traditional spectroscopy.

As described in Section \ref{sec:slam}, the worst regression model, i.e., the constant model, has $\rm MSE=1$ in standardized space. Any better model should reduce the MSE of this pixel to a value far below 1.
We denote this MSE as $\rm MSE_{full}$. Then, $\rm 1-MSE_{full}$ can be considered as a proper measure of the fraction of the variation of the pixel being \textit{explained} by model.
We define $\rm 1-MSE_{full}$ as the full coefficient of dependency (the full COD) of stellar labels, i.e.
\begin{equation}
\rm COD_{full} = 1-MSE_{full}.
\end{equation}
The maximum and minimum value of COD$_{\rm full}$ are 1 and 0, respectively. The larger COD$_{\rm full}$ is, the better the model is. 

Let \bz{$\boldsymbol{L}$} denote the collection of stellar labels (\teff, \logg\ and etc.), and let $l$ denote one specific stellar label in \bz{$\boldsymbol{L}$}. To derive the contribution of each stellar labels in COD$_{\rm full}$, we did a Leave-One-Label-Out training.
For example, to quantify the contribution of $l$, we \bz{remove} $l$ from \bz{$\boldsymbol{L}$} and train SVR on \bz{the other stellar labels}.
We write the obtained MSE in this case as MSE$_l$.
In principle, MSE$_l$ equals to or is larger than MSE$_{\rm full}$ because the model ignores the variation of the spectra driven by the stellar label $l$. The difference, ${\rm MSE}_l - {\rm MSE}_{\rm full}$,  measures the loss due to excluding stellar label $l$ in the model.
We then define COD$_l$ as% this to be the fraction of $\rm 1-MSE_{full}$ contributed by stellar label $l$.
%Therefore we have
\begin{equation}
{\rm COD}(l) = {\rm COD}_{\rm full}  %\left(1-MSE_{full}\right)
\times \frac{{\rm MSE}_{l} - {\rm MSE}_{\rm full}}{\sum_{l \in \boldsymbol{L}} \left({\rm MSE}_{l} - {\rm MSE}_{\rm full}\right)}.
\end{equation}
%where Qs represents the set of all stellar labels and Q represents a specified one among Qs.
%For example, we leave \teff\ aside and model spectra on only \logg\ and \feh and obtained $\rm{MSE}_{T_{\rm eff}}$.
By definition, $\sum_l {\rm COD}(l) = {\rm COD}_{\rm full}$.

We derive the CODs of \teff, \logg\ and \feh\ for the training samples used in Section \ref{sec:test} (covering \teff\ from 4000 to 8000\,K) and show them in Figure~\ref{fig_lamost_cods}. For most part of the spectra, the COD spectra are amazingly consistent with the empirical knowledge about which spectral lines are sensitive to which stellar labels. The blue, orange, and green filled regions represent for the COD(\teff), COD(\logg), and COD(\feh), respectively. The most significant features are at around the Balmer lines. At H$\delta$, H$\gamma$, and H$\beta$, COD(\teff) is very large and dominant, while COD(\feh) and COD(\logg) are small, meaning that these pixels depend mainly on \teff\ rather than \logg\ and \feh.
Across the whole spectrum, the Balmer lines are the most prominent features sensitive to effective temperature.
% Therefore, \teff\ values are mainly derived with these features.
The line centers of Balmer lines appears to be slightly different than the line wings, which reflects a different mechanism in formation of line centers. 

\bz{
From Figure~\ref{fig_lamost_cods}, it seems that the most information of \logg\ (for K giant stars) comes from the $\sim$4200 $\rm \AA$ region and the Mg I triplet at around $\lambda$5175. We can find that most of the dependence on \logg\ come from the doublets, triplets and line wings. The pixels located at the wings of the three lines of Mg I triplet show high dependence on \logg. This behavior is largely different compared to the COD(\teff) and COD(\logg).
}

The COD(\feh) is largely coincident with the positions of metal lines such as the Fe $\lambda$5709 and Fe $\lambda$5782. In our experience, the Ca II K and H lines are good proxies of metallicity.
However, because the inverse variance of the LAMOST spectra in the very blue part of the spectrum ($\lambda \sim 3936$ $\rm{\AA}$ and $3970$ $\rm \AA$) are frequently marked as bad pixels, many of the Ca II H and K lines are unavailable. Therefore, the COD(\feh) does not show strong dependence at the Ca II H an K lines.
% The whole part of the spectrum ($\lambda < 5800 \rm{\AA}$) gives good constraint in \feh.

The picture gives us a good interpretation of how machine learning algorithms \textit{learn from the data} and help human \textit{understand the data}.
Although the CODs are very similar to the gradient $\left( \dfrac{\partial f}{\partial l} \right)$ which is also shown in other works such as \cite{Ness2015}, they are different. 
The gradient is essentially the first-order partial derivative, so it reflects the \textit{local} dependence of the fluxes on stellar labels only in the first-order.
In contrast, the CODs measure the \textit{global} dependence and do not rely on the specific analytic models to map the stellar labels to the spectral fluxes.

%___________________________________________________________________________________
%
%               Conclusion
%___________________________________________________________________________________
%

\section{Conclusions}
\label{sec:conclusion}
\bz{Following the idea of the data-driven methods, we} present the Steller LAbel Machine (\slam), an SVR-base method, in this work. Taking advantages of the non-parametric nature of SVR, \slam\ is able to fit multi-dimensional and highly non-linear relationship between the fluxes and stellar labels, which is very different from \cannon.

We validate our method with LAMOST DR5 to investigate the performance and precision of the predicted labels. The CV scatters of \teff, \logg, and \feh\ at high SNR$_g$ ($\sim$100) are 50 K, 0.09 dex, and 0.07 dex, respectively.

We also use our method to predict stellar labels of LAMOST DR5 K giant stars with the training labels from APOGEE DR15. The performance assessment indicates that \slam\ is moderately better than \cannon. The CV scatters at high SNR$-g$ end are 49 K, 0.10 dex, 0.037 dex, 0.026 dex, 0.058 dex, and 0.106 dex for \teff, \logg, \mh, \am, \cm, and \nm, respectively. %These results are quite fascinating and encouraging although not perfect.
%With these stellar labels, we acquire more than a million red giant stars which are of interest by many astronomers.
We provide a downloadable catalog composed of \slam-derived \teff, \logg, \mh, \am, \cm, and \nm\ for more than a million LAMOST K giant stars.

\begin{table*}
	\centering
	\caption{An example of the catalog of the LAMOST DR5 K giant stars with \slam-derived stellar labels. Column 1 is the LAMOST IDs of the objects, Column 2-3 are the sky coordinates of the objects, Column 4-9 are the \slam-predicted stellar labels and Column 10-15 are the corresponding errors, Column 16 is the convergence flag of estimated stellar labels (successfully converged if True), Column 17 is the root mean squared deviation between the observed and fitted spectra, Column 18 is the index of our selection of K giant stars, Column 19 is the label-distances and Column 20 is the APOGEE observation flag (observed by APOGEE if True). }
	\label{tab:example_table}

\begin{tabular}{ccccccccc}
\hline\hline
LAMOST obsid & R.A.(J2000) & Dec.(J2000) & \teff\ & \logg\ & \mh\ & \am\ & \cm\ & \nm\ \\
\hline
 & (deg) & (deg) & (K) & (dex) & (dex) & (dex) & (dex) & (dex) \\
\hline
101001 & 332.20227 & -2.05677 & 4305.1 & 3.04 & -0.5260 & 0.2352 & -0.0115 & 0.2408 \\
101002 & 332.47158 & -2.08501 & 3920.2 & 1.04 & -0.5369 & 0.2168 & 0.1311 & 0.1296 \\
101004 & 332.43155 & -2.06237 & 4590.2 & 2.60 & -0.2077 & 0.1938 & 0.1108 & 0.1723 \\
101005 & 332.53546 & -2.11644 & 5016.2 & 3.40 & -0.6075 & 0.0870 & 0.0027 & -0.3379 \\
101006 & 332.34278 & -1.91919 & 5132.6 & 3.04 & -1.6101 & 0.3239 & -0.1626 & -0.2507 \\
101007 & 332.51256 & -1.84141 & 4460.6 & 1.29 & -1.6390 & 0.3090 & -0.2499 & 0.3238 \\
101008 & 332.36874 & -1.95577 & 5262.5 & 3.30 & 0.0350 & 0.2063 & -0.0615 & 0.2708 \\
101009 & 332.20666 & -1.86865 & 5672.1 & 4.05 & -0.3786 & 0.1140 & -0.1829 & 0.2263 \\
101010 & 332.39213 & -1.86556 & 4460.6 & 1.29 & -1.6390 & 0.3090 & -0.2499 & 0.3238 \\
101013 & 332.32152 & -2.12202 & 4849.3 & 1.92 & -2.1013 & 0.3850 & -0.0635 & -0.7601 \\
\hline
\end{tabular}

\begin{tabular}{ccccccccccc}
\hline\hline
$\sigma$(\teff) & $\sigma$(\logg) & $\sigma$(\mh) & $\sigma$(\am) & $\sigma$(\cm) & $\sigma$(\nm) & convergence & rmse & Kgiant(cut) & D(Ho2017) & in APOGEE\\
\hline
(dex) & (dex) & (dex) & (dex) & (dex) & (dex) & bool &  & bool & & bool\\
\hline
143.3 & 0.35 & 0.2103 & 0.0690 & 0.1313 & 0.1763 & False & 0.096 & False & 113.34 & False \\
205.2 & 0.54 & 0.3586 & 0.0940 & 0.1698 & 0.2072 & False & 0.158 & False &  & False \\
241.6 & 0.65 & 0.4539 & 0.1080 & 0.1907 & 0.2230 & False & 0.950 & False &  & False \\
174.4 & 0.44 & 0.2825 & 0.0818 & 0.1512 & 0.1927 & False & 0.127 & False &  & False \\
226.2 & 0.61 & 0.4130 & 0.1021 & 0.1820 & 0.2165 & False & 0.487 & False &  & False \\
241.3 & 0.65 & 0.4530 & 0.1079 & 0.1905 & 0.2228 & False & 1.575 & False &  & False \\
103.2 & 0.23 & 0.1264 & 0.0518 & 0.1033 & 0.1517 & True & 0.068 & True & 39.26 & False \\
139.1 & 0.34 & 0.2011 & 0.0673 & 0.1285 & 0.1740 & True & 0.077 & False & 1.43 & False \\
236.1 & 0.64 & 0.4391 & 0.1059 & 0.1876 & 0.2206 & False & 1.022 & False &  & False \\
188.6 & 0.49 & 0.3170 & 0.0874 & 0.1599 & 0.1995 & False & 0.187 & False &  & False \\
\hline
\end{tabular}
\end{table*}

\section*{Acknowledgements}

This work is supported by the National Natural Science Foundation of China (NSFC) with grant No. 11835057.

% LAMOST
Guoshoujing Telescope (the Large Sky Area Multi-Object Fiber Spectroscopic Telescope LAMOST) is a National Major Scientific Project built by the Chinese Academy of Sciences. Funding for the project has been provided by the National Development and Reform Commission. LAMOST is operated and managed by the National Astronomical Observatories, Chinese Academy of Sciences.

% SDSS-IV
Funding for the Sloan Digital Sky Survey IV has been provided by the Alfred P. Sloan Foundation, the U.S. Department of Energy Office of Science, and the Participating Institutions. SDSS-IV acknowledges
support and resources from the Center for High-Performance Computing at
the University of Utah. The SDSS web site is www.sdss.org.

SDSS-IV is managed by the Astrophysical Research Consortium for the 
Participating Institutions of the SDSS Collaboration including the 
Brazilian Participation Group, the Carnegie Institution for Science, 
Carnegie Mellon University, the Chilean Participation Group, the French Participation Group, Harvard-Smithsonian Center for Astrophysics, 
Instituto de Astrof\'isica de Canarias, The Johns Hopkins University, Kavli Institute for the Physics and Mathematics of the Universe (IPMU) / 
University of Tokyo, the Korean Participation Group, Lawrence Berkeley National Laboratory, 
Leibniz Institut f\"ur Astrophysik Potsdam (AIP),  
Max-Planck-Institut f\"ur Astronomie (MPIA Heidelberg), 
Max-Planck-Institut f\"ur Astrophysik (MPA Garching), 
Max-Planck-Institut f\"ur Extraterrestrische Physik (MPE), 
National Astronomical Observatories of China, New Mexico State University, 
New York University, University of Notre Dame, 
Observat\'ario Nacional / MCTI, The Ohio State University, 
Pennsylvania State University, Shanghai Astronomical Observatory, 
United Kingdom Participation Group,
Universidad Nacional Aut\'onoma de M\'exico, University of Arizona, 
University of Colorado Boulder, University of Oxford, University of Portsmouth, 
University of Utah, University of Virginia, University of Washington, University of Wisconsin, 
Vanderbilt University, and Yale University.


\vspace{5mm}
%\facilities{LAMOST, SDSS/APOGEE}


\software{scikit-learn \citep{2012arXiv1201.0490P},
astropy \citep{2018AJ....156..123A, 2013A&A...558A..33A},  
IPython \citep{2007CSE.....9c..21P}, 
Scipy\citep{jones2001}}



\appendix
\section{How to choose the best hyper-parameters }
\label{app:a}
\begin{figure*}
\plotone{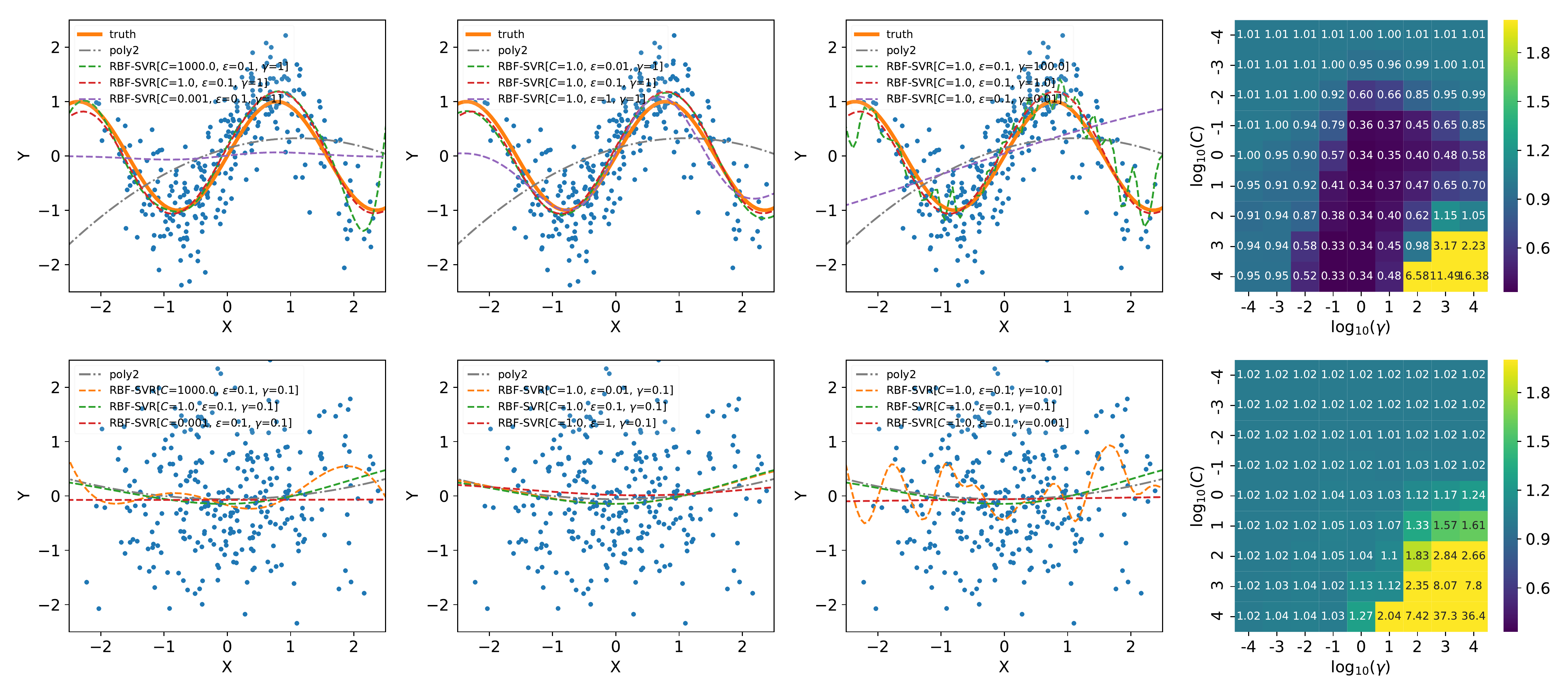}
\caption{Two examples of how the best hyper-parameters are chosen for SVR.
\label{fig_hyperparameters}}
\end{figure*}

Basically there are two kinds of pixels, i.e., spectral line pixels and continuum pixels.
The former kind contains much information of stellar labels while the latter contains almost no information.
In this section, we show how the best hyper-parameters are chosen in these two cases.

We simulate the first case in the upper row of Figure \ref{fig_hyperparameters}.
We use $\mathcal{N}(\mu, \sigma^2)$ to denote the normal distribution with mean of $\mu$ and variance of $\sigma^2$.
The $x$ data follows $\mathcal{N}(0, 1)$ and $y=\sin{2x}+noise$, where the $noise$ here follows $\mathcal{N}(0, 0.16)$.
In the first three panels of the upper row, we show how the fitting performance changes when varying one hyper-parameter, i.e., $C$, $\epsilon$, and $\gamma$, respectively.
We also superposed a quadratic model in gray dashed dotted line.
We can infer from these three panels that $C$ (the penalty level) and $\gamma$ (the width of the gaussian kernel or the softness of the SVR model) are more important relative to $\epsilon$ (the tube radius).
In the fourth panel, we show the color-coded 10-fold CV MSE as a function of $C$ and $\gamma$.
At $\log{\gamma}\sim0$ and $0<\log{C}<4$, the CV MSE reaches the mimimum.
At $\log{\gamma}\sim4$ and $\log{C}\sim4$, where the SVR has a high penalty for outliers and is extremely soft, the CV MSE is even larger than 1, which means \textit{over-fitting} occurs.
Clearly, we are able to determine the best set of hyper-parameters by choosing the one with lowest CV MSE in this diagram.

%The point here is that the MSE we used in this example is 10-fold cross-validated MSE, which is the average 10 MSEs from 10-fold cross-validation, therefore it comes close to the test error but not training error.

To simulate the latter case, we make both $x$ and $y$ follow $\mathcal{N}(0, 1)$.
In the lower row of Figure \ref{fig_hyperparameters}, we show similar plots.
In this case, it is seen in the last panel that the CV MSE is around 1 for most of the combinations of hyper-parameters.
And again \textit{over-fitting} arises at large $C$ and large $\gamma$.
\slam\ chooses the set of hyper-parameters with the lowest CV MSE, which prevents the model from \textit{over-fitting}.
%In the first three panels, we can see that the SVR model with CV MSE $\sim1$ are very simple.

\section{The source code of \slam}
\bz{
The source code of \slam\ is available on GitHub \url{https://github.com/hypergravity/astroslam} under an MIT License and the latest version is archived in Zenodo \citep{zhang2019}.
It can also be directly installed by running the following command in a terminal, {\ttfamily pip install astroslam}.
}


\begin{thebibliography}{}
\bibitem[\protect\citeauthoryear{Albareti et al.}{2017}]{Albareti2017}Albareti F.~D., et al., 2017, ApJS, 233, 25 
%\bibitem[\protect\citeauthoryear{Aguado et al.}{2019}]{2019ApJS..240...23A} Aguado, D.~S., Ahumada, R., Almeida, A., et al.\ 2019, \apjs, 240, 23
\bibitem[\protect\citeauthoryear{Astropy Collaboration et al.}{2013}]{2013A&A...558A..33A} Astropy Collaboration, Robitaille, T.~P., Tollerud, E.~J., et al.\ 2013, \aap, 558, A33 
\bibitem[\protect\citeauthoryear{Astropy Collaboration et al.}{2018}]{2018AJ....156..123A} Astropy Collaboration, Price-Whelan, A.~M., Sip{\H o}cz, B.~M., et al.\ 2018, \aj, 156, 123 
\bibitem[\protect\citeauthoryear{Bailer-Jones}{1997}]{BailerJones1997}Bailer-Jones C.~A.~L., 1997, PASP, 109, 932 
\bibitem[\protect\citeauthoryear{Bailer-Jones et al.}{1997}]{BailerJones1997}Bailer-Jones C.~A.~L., Irwin M., Gilmore G., von Hippel T., 1997, MNRAS, 292, 157 
\bibitem[\protect\citeauthoryear{Bailer-Jones, Irwin, \& von Hippel}{1998}]{BailerJones1998}Bailer-Jones C.~A.~L., Irwin M., von Hippel T., 1998, MNRAS, 298, 361 
\bibitem[\protect\citeauthoryear{Bailer-Jones}{2000}]{BailerJones2000}Bailer-Jones C.~A.~L., 2000, A\&A, 357, 197 
\bibitem[\protect\citeauthoryear{Beers et al.}{2006}]{Beers2006}Beers T.~C., et al., 2006, MmSAI, 77, 1171 
\bibitem[\protect\citeauthoryear{Bergemann et al.}{2016}]{Bergemann2016}Bergemann M., et al., 2016, A\&A, 594, A120 
\bibitem[\protect\citeauthoryear{Blanco-Cuaresma et al.}{2016}]{BlancoCuaresma2016}Blanco-Cuaresma S., et al., 2016, csss.conf, 22 
\bibitem[Buder et al.(2018)]{2018MNRAS.478.4513B} Buder, S., Asplund, M., Duong, L., et al.\ 2018, \mnras, 478, 4513
\bibitem[\protect\citeauthoryear{Brahm et al.}{2017}]{Brahm2017}Brahm R., Jord{\'a}n A., Hartman J., Bakos G., 2017, MNRAS, 467, 971 
\bibitem[Bu, \& Pan(2015)]{2015MNRAS.447..256B} Bu, Y., \& Pan, J.\ 2015, \mnras, 447, 256
\bibitem[\protect\citeauthoryear{Casey et al.}{2016}]{Casey2016}Casey A.~R., Hogg D.~W., Ness M., Rix H.-W., Ho A.~Q., Gilmore G., 2016, arXiv, arXiv:1603.03040 
\bibitem[\protect\citeauthoryear{Casey et al.}{2017}]{Casey2017}Casey A.~R., et al., 2017, ApJ, 840, 59 
\bibitem[\protect\citeauthoryear{Castelli, \& Kurucz}{2003}]{2003IAUS..210P.A20C} Castelli, F., \& Kurucz, R.~L.\ 2003, Modelling of Stellar Atmospheres, A20
\bibitem[\protect\citeauthoryear{Chang \& Lin}{2011}]{Chang2011} Chang., C.-C., Lin., C.-J., 2011, ACM Transactions on Intelligent Systems and Technology, 2:27:1
\bibitem[\protect\citeauthoryear{Cui et al.}{2012}]{2012RAA....12.1197C} Cui X.-Q., et al., 2012, RAA, 12, 1197 
\bibitem[\protect\citeauthoryear{de Boor}{1978}]{1978pgts.book.....D} de Boor, C.\ 1978, Applied Mathematical Sciences
\bibitem[\protect\citeauthoryear{Deng et al.}{2012}]{2012RAA....12..735D} Deng, L.-C., Newberg, H.~J., Liu, C., et al.\ 2012, Research in Astronomy and Astrophysics, 12, 735
\bibitem[\protect\citeauthoryear{Freeman}{2012}]{Freeman2012}Freeman K.~C., 2012, ASPC, 458, 393 
\bibitem[Garc{\'\i}a P{\'e}rez et al.(2016)]{2016AJ....151..144G} Garc{\'\i}a P{\'e}rez, A.~E., Allende Prieto, C., Holtzman, J.~A., et al.\ 2016, \aj, 151, 144
\bibitem[\protect\citeauthoryear{Gieseke et al.}{2017}]{Gieseke2017} Gieseke F., et al., 2017, MNRAS, 472, 3101 
\bibitem[\protect\citeauthoryear{Gilmore et al.}{2012}]{Gilmore2012}Gilmore G., et al., 2012, Msngr, 147, 25 
\bibitem[\protect\citeauthoryear{Ho et al.}{2017a}]{Ho2017a}Ho A.~Y.~Q., et al., 2017a, ApJ, 836, 5 
\bibitem[\protect\citeauthoryear{Ho et al.}{2017b}]{Ho2017b}Ho A.~Y.~Q., Rix H.-W., Ness M.~K., Hogg D.~W., Liu C., Ting Y.-S., 2017b, ApJ, 841, 40 
\bibitem[\protect\citeauthoryear{Hogg et al.}{2016}]{Hogg2016}Hogg D.~W., et al., 2016, ApJ, 833, 262 
\bibitem[Holtzman et al.(2018)]{2018AJ....156..125H} Holtzman, J.~A., Hasselquist, S., Shetrone, M., et al.\ 2018, \aj, 156, 125
\bibitem[\protect\citeauthoryear{Husser et al.}{2013}]{Husser2013}Husser T.-O., Wende-von Berg S., Dreizler S., Homeier D., Reiners A., Barman T., Hauschildt P.~H., 2013, A\&A, 553, A6 
\bibitem[Jones et al.(2001)]{jones2001} Jones E, Oliphant E, Peterson P, et al. SciPy: Open Source Scientific Tools for Python, 2001, http://www.scipy.org/
\bibitem[Koleva et al.(2009)]{2009A&A...501.1269K} Koleva, M., Prugniel, P., Bouchard, A., et al.\ 2009, \aap, 501, 1269
\bibitem[\protect\citeauthoryear{Kuntzer, Tewes, \& Courbin}{2016}]{Kuntzer2016} Kuntzer T., Tewes M., Courbin F., 2016, A\&A, 591, A54 
\bibitem[\protect\citeauthoryear{Lee et al.}{2011}]{Lee2011}Lee Y.~S., et al., 2011, AJ, 141, 90 
\bibitem[\protect\citeauthoryear{Lee et al.}{2008}]{Lee2008}Lee Y.~S., et al., 2008, AJ, 136, 2022 
\bibitem[\protect\citeauthoryear{Li, Pan, \& Duan}{2017}]{Li2017}Li X.-R., Pan R.-Y., Duan F.-Q., 2017, RAA, 17, 036 
\bibitem[Li et al.(2014)]{2014ApJ...790..105L} Li, X., Wu, Q.~M.~J., Luo, A., et al.\ 2014, \apj, 790, 105
\bibitem[Liu et al.(2014)]{2014RAA....14..423L} Liu, C.-X., Zhang, P.-A., \& Lu, Y.\ 2014, Research in Astronomy and Astrophysics, 14, 423-432
\bibitem[Lu, \& Li(2015)]{2015MNRAS.452.1394L} Lu, Y., \& Li, X.\ 2015, \mnras, 452, 1394
\bibitem[\protect\citeauthoryear{Liu et al.}{2012}]{Liu2012}Liu C., Bailer-Jones C.~A.~L., Sordo R., Vallenari A., Borrachero R., Luri X., Sartoretti P., 2012, MNRAS, 426, 2463 
\bibitem[Liu et al.(2014)]{Liu2014} Liu, C., Deng, L.-C., Carlin, J.~L., et al.\ 2014, \apj, 790, 110
\bibitem[Liu et al.(2015)]{Liu2015} Liu, C., Fang, M., Wu, Y., et al.\ 2015, \apj, 807, 4
%\bibitem[\protect\citeauthoryear{Liu et al.}{2015}]{Liu2015}Liu C., et al., 2015, RAA, 15, 1137 
\bibitem[\protect\citeauthoryear{Liu et al.}{2014}]{2014IAUS..298..310L} Liu X.-W., et al., 2014, IAUS, 298, 310 
\bibitem[\protect\citeauthoryear{Liu, Zhao, \& Hou}{2015}]{2015RAA....15.1089L} Liu X.-W., Zhao G., Hou J.-L., 2015, RAA, 15, 1089 
\bibitem[Liu et al.(2019)]{Liu2019} Liu, Z., Cui, W., Liu, C., et al.\ 2019, \apjs, 241, 32
\bibitem[\protect\citeauthoryear{Majewski}{2012}]{Majewski2012}Majewski S.~R., 2012, AAS, 219, 205.06 
\bibitem[\protect\citeauthoryear{Majewski et al.}{2017}]{2017AJ....154...94M} Majewski, S.~R., Schiavon, R.~P., Frinchaboy, P.~M., et al.\ 2017, \aj, 154, 94 
\bibitem[\protect\citeauthoryear{Majewski, APOGEE Team, \& APOGEE-2 Team}{2016}]{Majewski2016}Majewski S.~R., APOGEE Team, APOGEE-2 Team, 2016, AN, 337, 863 
\bibitem[\protect\citeauthoryear{Maraston \& Str{\"o}mb{\"a}ck}{2011}]{Maraston2011}Maraston C., Str{\"o}mb{\"a}ck G., 2011, MNRAS, 418, 2785 
\bibitem[\protect\citeauthoryear{M{\'e}sz{\'a}ros et al.}{2013}]{Meszaros2013}M{\'e}sz{\'a}ros S., et al., 2013, AJ, 146, 133 
\bibitem[\protect\citeauthoryear{M{\'e}sz{\'a}ros \& Allende Prieto}{2013}]{Meszaros2013}M{\'e}sz{\'a}ros S., Allende Prieto C., 2013, MNRAS, 430, 3285 
\bibitem[Mor{\'e}(1978)]{1978LNM...630..105M} Mor{\'e}, J.~J.\ 1978, Lecture Notes in Mathematics, Berlin Springer Verlag, 105
\bibitem[\protect\citeauthoryear{Ness et al.}{2015}]{Ness2015}Ness M., Hogg D.~W., Rix H.-W., Ho A.~Y.~Q., Zasowski G., 2015, ApJ, 808, 16 
\bibitem[\protect\citeauthoryear{Ness et al.}{2016}]{Ness2016}Ness M., Hogg D.~W., Rix H.-W., Martig M., Pinsonneault M.~H., Ho A.~Y.~Q., 2016, ApJ, 823, 114 
\bibitem[\protect\citeauthoryear{Ness et al.}{2016}]{Ness2016yCat} Ness M., Hogg D.~W., Rix H.-W., Martig M., Pinsonneault M.~H., Ho A.~Y.~Q., 2016, yCat, 182,  
\bibitem[\protect\citeauthoryear{Newberg et al.}{2012}]{Newberg2012}Newberg H.~J., et al., 2012, ASPC, 458, 405 
\bibitem[\protect\citeauthoryear{Pancino et al.}{2017}]{Pancino2017}Pancino E., et al., 2017, A\&A, 598, A5 
\bibitem[\protect\citeauthoryear{Pedregosa et al.}{2012}]{Pedregosa2012}Pedregosa F., et al., 2012, arXiv, arXiv:1201.0490 
\bibitem[\protect\citeauthoryear{Pedregosa et al.}{2012}]{2012arXiv1201.0490P} Pedregosa, F., Varoquaux, G., Gramfort, A., et al.\ 2012, arXiv e-prints, arXiv:1201.0490
\bibitem[\protect\citeauthoryear{Perez, \& Granger}{2007}]{2007CSE.....9c..21P} Perez, F., \& Granger, B.~E.\ 2007, Computing in Science and Engineering, 9, 21
\bibitem[Prugniel et al.(2007)]{2007astro.ph..3658P} Prugniel, P., Soubiran, C., Koleva, M., et al.\ 2007, arXiv e-prints, astro-ph/0703658
\bibitem[\protect\citeauthoryear{Rix et al.}{2016}]{2016ApJ...826L..25R} Rix, H.-W., Ting, Y.-S., Conroy, C., et al.\ 2016, \apj, 826, L25.
\bibitem[\protect\citeauthoryear{Skrutskie et al.}{2006}]{2006AJ....131.1163S} Skrutskie M.~F., et al., 2006, AJ, 131, 1163 
\bibitem[\protect\citeauthoryear{Smola \& Sch{\"o}lkopf}{2004}]{ss2004} Smola, A. J., Sch{\"o}lkopf, B, Statistics and Computing, 14, 199
\bibitem[\protect\citeauthoryear{Soubiran et al.}{2010}]{Soubiran2010}Soubiran C., Le Campion J.-F., Cayrel de Strobel G., Caillo A., 2010, A\&A, 515, A111 
\bibitem[\protect\citeauthoryear{Steinmetz et al.}{2006}]{Steinmetz2006}Steinmetz M., et al., 2006, AJ, 132, 1645 
\bibitem[\protect\citeauthoryear{Ting, Conroy, \& Rix}{2016}]{Ting2016}Ting Y.-S., Conroy C., Rix H.-W., 2016, ApJ, 826, 83 
%measure 14 elements
\bibitem[\protect\citeauthoryear{Ting et al.}{2017}]{Ting2017}Ting Y.-S., Rix H.-W., Conroy C., Ho A.~Y.~Q., Lin J., 2017, ApJ, 849, L9 
\bibitem[\protect\citeauthoryear{Ting et al.}{2017}]{2017ApJ...843...32T} Ting Y.-S., Conroy C., Rix H.-W., Cargile P., 2017, ApJ, 843, 32
\bibitem[Ting et al.(2019)]{2019ApJ...879...69T} Ting, Y.-S., Conroy, C., Rix, H.-W., et al.\ 2019, \apj, 879, 69
%\bibitem[\protect\citeauthoryear{Ting et al.}{2018}]{2018arXiv180401530T} Ting Y.-S., Conroy C., Rix H.-W., Cargile P., 2018, arXiv e-prints, arXiv:1804.01530
\bibitem[\protect\citeauthoryear{Tonry et al.}{2012}]{2012ApJ...750...99T} Tonry J.~L., et al., 2012, ApJ, 750, 99 
\bibitem[\protect\citeauthoryear{Wu et al.}{2011}]{2011RAA....11..924W} Wu Y., et al., 2011, RAA, 11, 924 
\bibitem[\protect\citeauthoryear{Wu et al.}{2014}]{2014IAUS..306..340W} Wu Y., Du B., Luo A., Zhao Y., Yuan H., 2014, IAUS, 306, 340 
\bibitem[\protect\citeauthoryear{Xiang et al.}{2015}]{Xiang2015} Xiang M.~S., et al., 2015, MNRAS, 448, 822 
\bibitem[\protect\citeauthoryear{Xiang et al.}{2017}]{Xiang2017}Xiang M.-S., et al., 2017, MNRAS, 464, 3657 
\bibitem[\protect\citeauthoryear{York et al.}{2000}]{2000AJ....120.1579Y} York D.~G., et al., 2000, AJ, 120, 1579 
\bibitem[\protect\citeauthoryear{Yuan et al.}{2015}]{2015MNRAS.448..855Y} Yuan H.-B., et al., 2015, MNRAS, 448, 855 
\bibitem[\protect\citeauthoryear{Zacharias et al.}{2013}]{2013AJ....145...44Z} Zacharias N., Finch C.~T., Girard T.~M., Henden A., Bartlett J.~L., Monet D.~G., Zacharias M.~I., 2013, AJ, 145, 44 
%\bibitem[Zhang(2019)]{zhang2019} Zhang, B.\ 2019, SLAM: Stellar LAbel Machine v1.2019.1005.0, Zenodo, doi:10.5281/zenodo.3473891
\bibitem[Zhang(2019)]{zhang2019} Zhang, B.\ 2019, SLAM: Stellar LAbel Machine v1.2019.1005.0, Zenodo, doi:10.5281/zenodo.3461503
\bibitem[Zhong et al.(2019)]{Zhong2019} Zhong, J., Li, J., Carlin, J.~L., et al.\ 2019, arXiv e-prints, arXiv:1908.01128

% \bibitem[Casey(2016)]{2016ApJS..223....8C} Casey, A.~R.\ 2016, \apjs, 223, 8
%\bibitem[J{\"o}nsson et al.(2018)]{2018AJ....156..126J} J{\"o}nsson, H., Allende Prieto, C., Holtzman, J.~A., et al.\ 2018, \aj, 156, 126
\end{thebibliography}
\end{document}